\documentclass{aa}  
\usepackage{graphicx}
\usepackage{txfonts}
\usepackage{hyperref}

\begin{document} 

   \title{Carbon-rich (DQ) white dwarfs in the Sloan Digital Sky
     Survey}
   \author{D. Koester
          \inst{1}
          \and
          S.O. Kepler\inst{2}
          }
     \institute{Institut f\"ur Theoretische
     Physik und Astrophysik, Universit\"at Kiel, 24098 Kiel,
     Germany\\ 
     \email{koester@astrophysik.uni-kiel.de} 
     \and Instituto de Fisica, Universidade Federal do Rio Grande do Sul, 
     91501-900 Porto-Alegre, RS, Brazil
     }

   \date{June 27, 2019}

\abstract
{Among the spectroscopically identified white dwarfs, a fraction
  smaller than 2\% have spectra dominated by carbon lines, mainly
  molecular C$_2$, but also in a smaller group by \ion{C}{i} and
  \ion{C}{ii} lines. These are together called DQ white dwarfs. }
{We want to derive atmospheric parameters \hbox{$T\sb{\rm eff}$},
  \hbox{$\log g$}, and carbon abundances for a large sample of these
  stars and discuss implications for their spectral evolution.}
{Sloan Digital Sky Survey spectra and $ugriz$ photometry were used,
  together with {\it Gaia} Data Release~2 parallaxes and $G$ band
  photometry. These were fitted to synthetic spectra and theoretical
  photometry derived from model atmospheres.}
%  results heading (mandatory)
{We found that the DQs hotter than \hbox{$T\sb{\rm
      eff}$}\ $\approx$10000\,K have masses $\approx$0.4\,
  \hbox{M$\sb{\odot}$}\ larger than the classical DQ, which have
  masses typical for the majority of white dwarfs
  ($\approx$0.6\,\hbox{M$\sb{\odot}$}). We found some evidence that
  the peculiar DQ  below 10000\,K also have significantly
  larger masses and may thus be the descendants of the hot and warm
  DQs above 10000\,K.  A significant fraction of the hotter objects
  with \hbox{$T\sb{\rm eff}$}\ $> 14500$\,K have atmospheres dominated
  by carbon.}
{}

   \maketitle
%
%-------------------------------------------------------------------

\section{Introduction}

White dwarfs of spectral type DQ are a subclass of hydrogen-poor
objects, i.e. hydrogen is not the dominant element as in the vast
majority of white dwarfs. Historically the first objects known
(henceforth simply called DQ) are characterized by Swan bands of the
C$_2$ molecule in the optical region. Later objects with strong
\ion{C}{i} resonance lines in the UV were added to this class
\citep{Sion.Greenstein.ea83}, even if they don't show any features in
the optical range. Systematic analyses of larger samples are
\citet{Dufour.Bergeron.ea05} and \citet{Koester.Knist06}.  Contrary to
the numerous metal-polluted hydrogen- and helium-rich white dwarfs,
where the metals are believed to be accreted from the debris of
disintegrating members of a former planetary system, the carbon in the
DQ (at least those below 10000\,K) is a result of dredge-up by an
extending convection zone in the upper helium layer
\citep{Koester.Weidemann.ea82, Pelletier.Fontaine.ea86}.

With the discovery of thousands of new white dwarfs thanks to the
Sloan Digital Sky Survey \citep[SDSS,][]{York.Adelman.ea00,
  Abolfathi.Aguado.ea18} a number of hotter objects with carbon
features were found
\citep{Liebert.Harris.ea03,Dufour.Fontaine.ea08}. The latter analysis
surprisingly concluded that the group of hot objects showing mostly
\ion{C}{ii} in the range of effective temperatures from 18000 to
24000\,K was in fact dominated by carbon in their atmospheres, with
only occasional traces of hydrogen, helium, or oxygen. Several members
of this group show magnetic fields, light variations, and unusually
high masses. In spite of many suggestions the evolutionary status of
these objects is not fully understood \citep{Fortier.Dufour15}.

Between the classical DQ and the new ``hot DQ'' (hDQ here) class
several objects with spectra dominated by \ion{C}{i} lines in the
optical range are now termed ``warm DQ'' (wDQ here)
\citep{Dufour.Vornanen.ea13,Fortier.Dufour15}.  These objects are
believed to be helium-dominated, but with large carbon contributions
of $\log$(N(C)/N(He)) (abbreviated as [C/He] henceforth) $\approx
-3$ to $-2$.

Several authors have found that the carbon abundances, when shown as a
function of \hbox{$T\sb{\rm eff}$}, have a clear tendency from higher
to lower values with decreasing temperature, and that a separate
second sequence exist with higher abundances \citep[see
  e.g.][]{dr12}. The upper sequence was also suggested to have higher
masses, although the data were very limited.

\begin{figure}
%\centering
\includegraphics[angle=-90,origin=c,width=9.0cm]{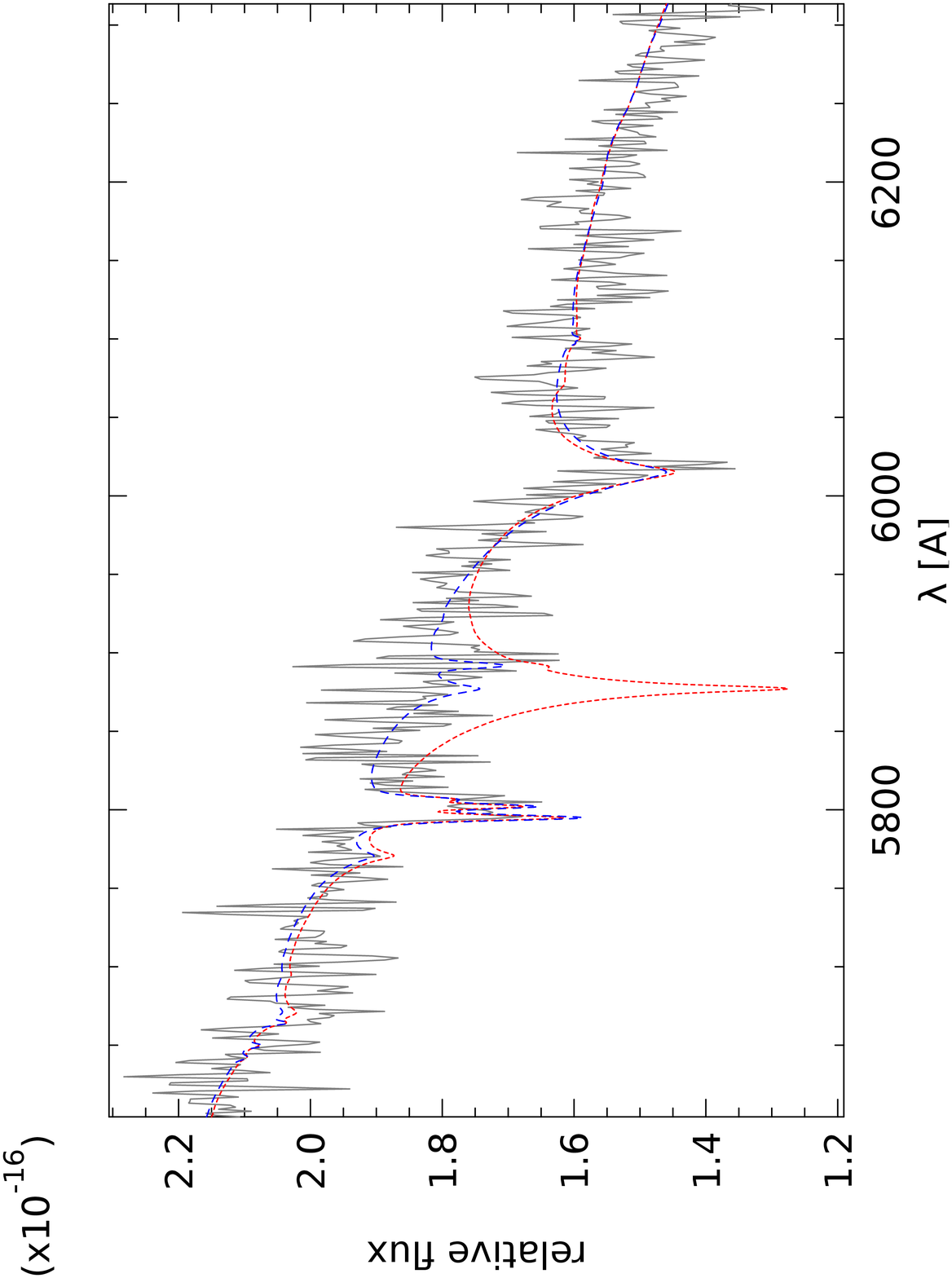}
\caption{SDSSJ1448+0519 (gray, continuous) compared with our models
  using the \citet{Coutu.Dufour.ea18} parameters (\hbox{$T\sb{\rm eff}$},
  \hbox{$\log g$},[C/He] = $15776,8.851,-1.70$, red dotted)
  and our own best fit (15966,8.943,0.300, blue dashed). Full SDSS
  names are given in Table~\ref{tabsample}}
\label{Figcoutu}
\end{figure}

\begin{figure}[h]
%\centering
\includegraphics[angle=-90,origin=c,width=9.0cm]{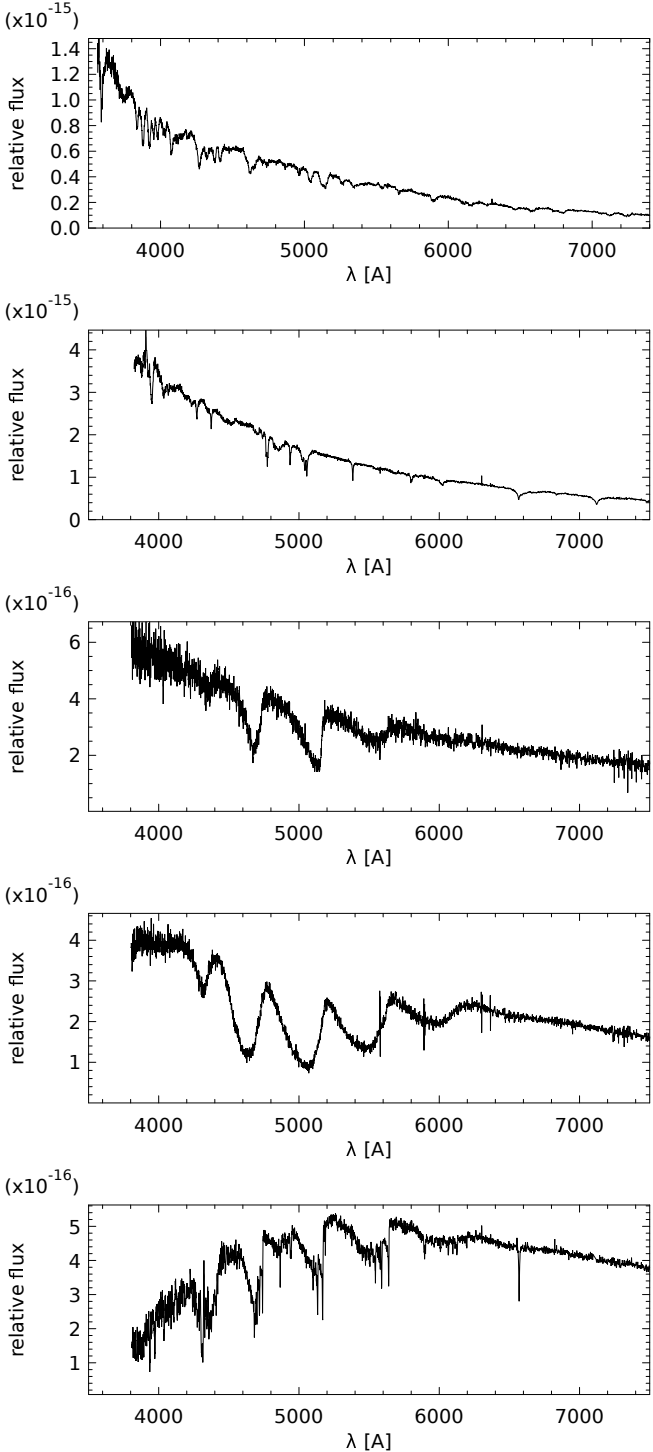}
\caption{Examples for the different carbon star types. From top to
  bottom (i) a hot DQ, hDQ (SDSSJ1104+2035), (ii) a warm DQ, wDQ
  (SDSSJ1728+5558), (iii) a classical DQ (SDSSJ1356$-$0009), (iv) a
  DQpec (SDSSJ2232$-$0744), and (v) a dwarf carbon star, dC
  (SDSSJ1127$-$0212). Note the extremely sharp band heads and Balmer
  lines.}
\label{Figtypes}
\end{figure}

\begin{table*}
  \caption{Observed data for the DQ + carbon star sample. p-m-f is the
    Plate-Epoch (MJD)-Fiber identification of the SDSS spectra,
    S/N the signal-to-noise ratio. $\pi$ is the {\it Gaia} parallax and G
    the {\it Gaia} magnitude. g is the SDSS g magnitude. The complete table
    is in the appendix.)
  }
\label{tabsample} 
\centering         
\begin{tabular}{ccrcccl}
\hline\hline
\noalign{\smallskip}
  SDSS~J      &            p-m-f   &      $\pi$ [msec] &  g   &     G  &   S/N& Type \\
  \hline
  \noalign{\smallskip}
000011.66$-$085008.3 &  7167-56604-0752 &   4.825 &  19.111 &  19.056 &  17.8 &    DQ\\
000705.02$+$282104.2 &  2824-54452-0602 &   5.420 &  19.702 &  19.590 &  13.6 &    DQ\\
001523.91$+$030909.0 &  4298-55511-0906 &   5.020 &  20.028 &  19.933 &   8.5 &    DQ\\
001908.63$+$184706.0 &  7592-56947-0123 &   6.594 &  19.123 &  19.160 &   9.7 &    wDQ\\
002531.50$-$110800.9 &  0653-52145-0086 &   9.255 &  17.999 &  17.949 &  20.5 &    DQ\\
003328.58$+$041834.6 &  4303-55508-0562 &   7.497 &  18.720 &  18.648 &  26.2 &    DQ\\
.....& & & & & & \\
\hline\\
\end{tabular}
\end{table*}

The surface gravities and masses are very difficult to determine
spectroscopically and only few were known until recently.  The
situation was much improved when the {\it Gaia} Data Release~2 (DR2)
parallaxes for many of these objects were published 
  \citep{GaiaCollaboration.Brown.ea18}.  Atmospheric parameters for a
  large sample of white dwarfs with {\it Gaia} parallaxes and SDSS
  spectra, which also included many DQ white dwarfs, was studied by
  \citet{GentileFusillo.Tremblay.ea19}.  The first application of
such a sample, which additionally obtained abundances, was
\citet{Coutu.Dufour.ea18}, which very clearly shows two separate
sequences in the mass distribution as well as the abundance
vs. \hbox{$T\sb{\rm eff}$}\ distribution.  All objects up to the
highest \hbox{$T\sb{\rm eff}$}\ of $\approx$15000\,K in their results
have helium dominated atmospheres with [C/He] $< -1.5$. The hottest
object is SDSSJ1448+0519 (we use this shorthand notation, the complete
official SDSS name can be found in Table~\ref{tabsample}, if necessary
via the p-m-f designation of the spectra), identified with the
parameters \hbox{$T\sb{\rm eff}$} = 15473\,K, \hbox{$\log g$} = 8.94,
[C/He] = $-1.70$. This object is also in our sample (described in
section \ref{sample}). We calculated a model with these
parameters. The model shows strong He lines at 4472, 5877, and
6679\,\AA\ (all these regions are not included in the figure on the
poster), which are completely absent in the observed spectrum. With
our analysis we find instead that this object has a carbon-dominated
atmosphere with an upper limit on the helium abundance of [He/C] <
$-0.30$. This raises the question of the nature of the transition
between hot and normal DQ, roughly in the temperature range
10000-18000\,K. Since the helium lines become invisible below
12000\,K, or even higher in low S/N spectra, this is a difficult
question, which is the major topic of this study. In addition we will
present some results for a fairly large sample of classical cool DQ
and a few hot DQ, which generally agree with previous findings.

%--------------------------------------------------------------------
\section{Observations and choice of sample \label{sample}}
The original sample in our study (572 stars) included all stars
classified as DQ from the 34973 white dwarfs in
\citet{dr7,dr10,dr12,dr14}, even if the classification was
uncertain. Therefore the DQs represent at most 1.6\% of the classified
white dwarfs from SDSS, and 9\% of the non-DAs.

From these we selected the spectra with highest S/N obtained by SDSS
up to DR14 for stars having {\it Gaia} DR2 parallaxes, necessary to estimate
the \hbox{$T\sb{\rm eff}$}, \hbox{$\log g$}, and mass.  We further
eliminated some wrong or uncertain type identifications, and objects
with low signal-to-noise ratio (S/N) or very weak features. We did not
aim at complete samples, but for a sample with accurate parameter
determinations. Most spectra have S/N $>10$, but we included some with
lower values if the features were strong and easily recognized. Also
eliminated were objects with {\it Gaia} parallax uncertainties larger than
$25\%$. We were then left with 304 stars, all showing clear carbon
features of Swan bands of the C$_2$ molecules, or lines from
\ion{C}{i} and/or \ion{C}{ii}. These stars are listed in
Table~\ref{tabsample}, together with some of the observed data.
\begin{table*}
\label{tabdq} 
\centering
\caption{Fit results for the classical DQ with Swan bands. The numbers
  in parentheses for \hbox{$T\sb{\rm eff}$}, \hbox{$\log g$}, [C/He],
  and $M$/\hbox{M$\sb{\odot}$} are the internal uncertainties.  The
  complete table is in the appendix.}
\begin{tabular}{cccccc}
\hline\hline
\noalign{\smallskip}
SDSS~J   &     p-m-f   &      Teff[K] & log g [cgs] &  [C/He]  &
$M$/\hbox{M$\sb{\odot}$}  \\
   \hline
   \noalign{\smallskip}
0000$-$0850 & 7167-56604-0752 & 7944 (117)& 7.453 (0.164)& $-5.172$ (1.369)& 0.311 (0.060)\\
0007$+$2821 & 2824-54452-0602 & 7529 ( 92)& 7.931 (0.142)& $-5.427$ (0.584)& 0.536 (0.082)\\
0015$+$0309 & 4298-55511-0906 & 7256 (132)& 8.004 (0.181)& $-5.678$ (2.117)& 0.578 (0.109)\\
0025$-$1108 & 0653-52145-0086 & 8402 ( 82)& 7.866 (0.041)& $-4.754$ (0.291)& 0.502 (0.023)\\
0033$+$0418 & 4303-55508-0562 & 7700 (108)& 7.847 (0.073)& $-5.792$ (0.601)& 0.489 (0.040)\\
.....& & & & & \\
\hline\\
\end{tabular}    
\end{table*}

\begin{table*}
  \caption{Comparison of average surface gravities for different types
    of helium-rich cool white dwarfs. Numbers < \hbox{$\log
      g$} > are averages in the \hbox{$T\sb{\rm eff}$}\ interval, slg
    the $1\sigma$ width of the distribution, not the uncertainties of
    the average. }
\label{tabcomparison} 
\centering         

\begin{tabular}{rrrrrrrrrr}
\hline\hline
\noalign{\smallskip}
\multicolumn{4}{c}{DQ} & \multicolumn{3}{c}{DZ}&\multicolumn{3}{c}{DB/DC}\\
\hbox{$T\sb{\rm eff}$}   &  <\hbox{$\log g$}>& slg & N&
<\hbox{$\log g$}>& slg&   N & <\hbox{$\log g$}>& slg&  N\\
      \hline
      \noalign{\smallskip}
   15000-12000&        &       &  0& 8.132&  0.168&  4& 8.016& 0.204& 15\\
   12000-11000&        &       &  0& 8.015&  0.113& 13& 8.007& 0.186& 39\\
   11000-10000&        &       &  0& 8.038&  0.116& 20& 7.979& 0.190&129\\
   10000- 9000&   7.938&  0.084&  9& 8.026&  0.106& 30& 8.027& 0.169&153\\
    9000- 8000&   7.921&  0.124& 85& 8.049&  0.127& 44& 8.040& 0.179&128\\
    8000- 7000&   7.941&  0.208& 81& 8.022&  0.126& 17& 8.085& 0.183&234\\
    7000- 6000&   7.961&  0.210& 38& 7.915&  0.151& 26& 8.039& 0.185&249\\
    6000- 5000&   7.886&  0.326&  8& 7.717&  0.153& 12&      &      &  0\\   
    total:    &   7.935&  0.184&221& 7.995&  0.158&166& 8.039& 0.185&947\\
\hline
\end{tabular}    
\end{table*}

\begin{table*}
\centering
\caption{Fit results for the warm DQ (wDQ). The second number in the
  columns for \hbox{$T\sb{\rm eff}$}, \hbox{$\log g$}, [C/He], and
  $M$/\hbox{M$\sb{\odot}$}\ are the internal uncertainties.}
\label{tabwdq} 
\begin{tabular}{cccccc}
\hline\hline
\noalign{\smallskip}
   SDSS~J   &     p-m-f   &      Teff[K] & log g [cgs] &  [C/He]  & $M$/\hbox{M$\sb{\odot}$} \\
   \hline
   \noalign{\smallskip}
0019$+$1847 & 7592-56947-0123 & 10280 (134) & 8.546 (0.075) & $-2.840$ (0.025) &  0.932 (0.046)\\
0236$+$2503 & 2399-53764-0059 & 14611 (308) & 8.777 (0.086) & $-1.823$ (0.057) &  1.067 (0.047)\\
0807$+$1949 & 4481-55630-0323 & 14593 (265) & 8.847 (0.071) & $ 1.044$ (0.119) &  1.105 (0.038)\\
0856$+$4513 & 7327-56715-0052 &  9353 (203) & 8.509 (0.159) & $-3.025$ (0.360) &  0.909 (0.099)\\
0859$+$3257 & 1272-52989-0309 &  9798 ( 59) & 8.516 (0.011) & $-2.958$ (0.002) &  0.914 (0.007)\\
0919$+$0236 & 3822-55544-0966 & 12447 (200) & 8.829 (0.069) & $-1.977$ (0.044) &  1.095 (0.038)\\
0936$+$0607 & 4871-55928-0024 & 12166 (292) & 8.816 (0.088) & $-2.210$ (0.046) &  1.088 (0.048)\\
0958$+$5853 & 5719-56014-0114 & 15444 (420) & 8.951 (0.054) & $-0.500$ (0.055) &  1.161 (0.029)\\
1049$+$1659 & 5352-56269-0732 & 13590 (305) & 8.995 (0.091) & $-0.757$ (0.203) &  1.186 (0.044)\\
1058$+$2846 & 2870-54534-0429 &  9742 (147) & 8.574 (0.074) & $-3.025$ (0.451) &  0.949 (0.044)\\
1100$+$1758 & 2485-54176-0119 & 12631 (244) & 8.756 (0.080) & $-1.913$ (0.064) &  1.055 (0.045)\\
1140$+$0735 & 5377-55957-0240 & 11407 (180) & 8.738 (0.000) & $-2.175$ (0.038) &  1.044 (0.040)\\
1140$+$1824 & 5891-56034-0946 &  9931 ( 87) & 8.431 (0.027) & $-3.000$ (0.014) &  0.860 (0.017)\\
1148$-$0126 & 0329-52056-0578 &  9847 ( 86) & 8.503 (0.019) & $-3.032$ (0.010) &  0.906 (0.012)\\
1203$+$6450 & 6975-56720-0446 & 12977 (133) & 8.786 (0.016) & $-1.909$ (0.023) &  1.071 (0.009)\\
1215$+$4700 & 6640-56385-0532 & 13940 (302) & 8.957 (0.056) & $-2.014$ (0.044) &  1.165 (0.027)\\ 
1331$+$3727 & 3984-55333-0118 & 16741 (334) & 9.028 (0.030) & $ 0.408$ (0.152) &  1.165 (0.054)\\
1332$+$2355 & 5995-56093-0918 & 15131 (356) & 8.779 (0.099) & $-0.750$ (0.218) &  1.068 (0.054)\\
1339$+$5036 & 6744-56399-0664 & 11680 (192) & 8.621 (0.063) & $-2.200$ (0.053) &  0.978 (0.037)\\
1341$+$0346 & 4786-55651-0184 & 13978 (422) & 8.834 (0.177) & $-2.151$ (0.049) &  1.098 (0.096)\\
1434$+$2258 & 6016-56073-0466 & 15750 (405) & 8.828 (0.084) & $ 0.725$ (0.727) &  1.094 (0.046)\\
1435$+$5318 & 1327-52781-0413 & 15658 (355) & 8.900 (0.063) & $-0.747$ (0.273) &  1.133 (0.034)\\
1448$+$0519 & 4858-55686-0082 & 15966 (388) & 8.943 (0.039) & $ 0.300$ (0.105) &  1.157 (0.022)\\
1622$+$1849 & 4060-55359-0346 & 16693 (479) & 9.129 (0.073) & $-0.079$ (0.164) &  1.157 (0.012)\\
1622$+$3004 & 4953-55749-0483 & 16131 (267) & 8.934 (0.022) & $-0.105$ (0.063) &  1.152 (0.012)\\
1728$+$5558 & 0358-51818-0296 & 14772 (193) & 8.869 (0.015) & $ 0.300$ (0.239) &  1.117 (0.008)\\
\hline\\
\end{tabular}    
\end{table*}

\begin{table*}
\centering
\caption{Fit results for the hot DQ (hDQ) with \hbox{$T\sb{\rm eff}$}
  $>$ 18000\,K.  The second number in the columns for \hbox{$T\sb{\rm eff}$},
  \hbox{$\log g$}, and $M$/\hbox{M$\sb{\odot}$}\ are the
  internal uncertainties. [C/He] = 2 was assumed for these fit (see
  text).}
\label{tabhdq} 
\begin{tabular}{rrrrrr}
\hline\hline
\noalign{\smallskip}
   SDSS~J   &     p-m-f   &      Teff[K] & log g [cgs] & $M$/\hbox{M$\sb{\odot}$} \\
   \hline
   \noalign{\smallskip}
0106$+$1513& 5131-55835-0736 & 23610  (107)&  8.572 (0.125) & 0.957 (0.073) \\
0818$+$0102& 2077-53846-0575 & 24483  ( 97)&  8.326 (0.055) & 0.807 (0.035) \\
1104$+$2035& 6428-56279-0282 & 25503  ( 60)&  8.661 (0.036) & 1.009 (0.020) \\
1200$+$2252& 2643-54208-0469 & 21880  (172)&  8.505 (0.243) & 0.915 (0.146) \\
1426$+$5752& 6803-56402-0916 & 18809  (161)&  8.722 (0.120) & 1.039 (0.066) \\
2200$-$0741& 0717-52468-0462 & 23097  ( 88)&  8.694 (0.057) & 1.026 (0.031) \\
2348$-$0942& 7166-56602-0536 & 22596  (119)&  8.652 (0.160) & 1.002 (0.091) \\
\hline\\
\end{tabular}    
\end{table*}

When we started the analysis, 29 objects indicated very low
surface gravities, at the lowest gravities of our model grids, but
still without a reasonable fit. Visual inspection revealed very narrow
bands and lines, in particular \hbox{H$\alpha$}. These objects are
very similar to those described in \citet{Whitehouse.Farihi.ea18}, and
we classify them as dwarf carbon stars (dC); they will not be
discussed further in this study.  Another 21 objects show Swan bands,
but of a peculiar, rounded shape and usually blue-shifted. This shift
is not really understood yet, but the discussions center on pressure
shifts of the Swan bands in high density helium \citet{Hall.Maxwell08,
  Kowalski10}. Since our models do not include this effect, we cannot
determine accurate parameters for these peculiar DQ (DQpec), but will
discuss their relevance for the spectral evolution of the DQ in
Sect.~\ref{discussion}.  This leaves 254 objects for the
analysis. Fig.~\ref{Figtypes} shows examples for all types discussed
here: DQ, warm DQ, hot DQ, peculiar DQ, and dwarf carbon stars dC.

\section{Theoretical models}
We calculated theoretical atmosphere models using basically the
methods described in \citet{Koester10}. However, the code was
completely rewritten, using the more modern FORTRAN95 language, which
supports a much stronger modularization of the programs. We updated
the equation of state (EOS) to use many more molecules, and include
non-ideal effects for neutral and charged interactions consistently in
EOS and secondary quantities as adiabatic derivatives, heat capacity,
and the adiabatic gradient. The absorption coefficients include more
negative ions than in previous versions of the code, and many
photoionization cross sections.

For the current DQ models 12 molecular species of H and C were
included, with 300 C$_2$ bands, 211 cross sections from \ion{C}{i} to
\ion{C}{iv}, and 713 spectral lines. We calculated two grids, the cool
one with 5000\,K $\leq$ \hbox{$T\sb{\rm eff}$}\ $\leq$ 15000\,K, step
250\,K, 7.0 $\leq$ \hbox{$\log g$}\ $\leq$ 9.5\, dex, step 0.25, $-9.5$
$\leq $[C/He] $\leq -2.0$, step 0.5, altogether 7216 models. The
second, hot grid had 9000\,K$\leq$ \hbox{$T\sb{\rm eff}$}\ $\leq$
20000\,K, step 250\,K, and 20000\,K $\leq$ \hbox{$T\sb{\rm eff}$}\
$\leq$ 30000\,K, step 1000\,K, 7.0 $\leq$
\hbox{$\log g$}\ $\leq$ 9.5\,dex, step 0.25, $-4.0 \leq$ [C/He]
$\leq$ +4.0, step 0.5, resulting in 10285 models. The total model
grids thus comprise 17501 models.

From the synthetic spectra we calculated theoretical photometry for
the SDSS $ugriz$ filters and the {\it Gaia} $G$ filter by convolving with
the filter bandpasses and applying appropriate zeropoints. We also
converted these theoretical magnitudes to absolute magnitudes by using
the Montreal
mass-radius-relation\footnote{\url{http://www.astro.umontreal.ca/~bergeron/CoolingModels}},
except for \hbox{$\log g$} $> 9.00$, where we used the 
zero-temperature relation for carbon from \citet{Hamada.Salpeter61}.
Since all objects in our sample have {\it Gaia} parallaxes, we can transform
the distance in a distance modulus $m-M =
5\log(\mathrm{distance/pc})-5$.  The advantage is that the distance is
expressed on a magnitude scale and can be used in the fitting in the
same way as the other 6 magnitudes ($ugrizG$).

\section{Analysis of photometric and spectroscopic data}
The new parallaxes provide a very strong constraint on the radius or
in our fitting method on the distance modulus $m-M$. With observed
magnitude $m$ and $M$ from our theoretical grid this is an equally
strong constraint of the surface gravity \hbox{$\log g$}. As this depends only
on the average level of the observed photometry and not on the shape of
the energy distribution, this is a much more robust determination than
the usual spectroscopic method using spectral line profiles.

Our tests indicated that the most successful procedure is an iteration
between spectroscopic and photometric fitting. For the cool and warm
DQ (\hbox{$T\sb{\rm eff}$} < 18000\,K) we used the photometry with
fixed [C/He] to obtain \hbox{$T\sb{\rm eff}$}\ and \hbox{$\log g$};
the spectroscopic fitting is used with fixed \hbox{$T\sb{\rm
    eff}$}\ and \hbox{$\log g$}\ and determines a new value for
[C/He]. The iteration stops when the latter value does not change
within the uncertainty. As a final consistency check we have done a
last spectroscopy fitting with \hbox{$\log g$}\ and [C/He] fixed, but
\hbox{$T\sb{\rm eff}$}\ freely variable. If this spectroscopic
\hbox{$T\sb{\rm eff}$}\ agreed with the photometric one, the procedure
was stopped.

For the hot DQ (hDQ) we used the photometry only for \hbox{$\log
  g$}\ and spectroscopy for \hbox{$T\sb{\rm eff}$}\ and [C/He]. This
worked better in the hotter temperature range, since the dependence of
the photometry on temperature gets smaller in the optical
Raleigh-Jeans range.  The absence of He lines in the hot DQ indicates
a [C/He] > 1. Since the spectra do not change significantly with even
higher ratios, we use a fixed [C/He]= 2 to obtain \hbox{$T\sb{\rm
    eff}$}\ from spectroscopy and \hbox{$\log g$}\ from photometry.

When using photometry for more distant objects, the interstellar
reddening is always a concern. Fortunately for our objects the
maximum reddening through the galactic disk, obtained from the SDSS
database is always small, mostly in the range 0.04 $\leq$ E(B-V)
$\leq$ 0.05, because the SDSS observations are by design mostly in a
direction perpendicular to the plane of the Galaxy. We have taken a
fraction of this value depending on the estimated distance, calculated
as in \citet{GentileFusillo.Tremblay.ea19}.
\begin{figure}
%\centering
\includegraphics[angle=-90,origin=c,width=8.0cm]{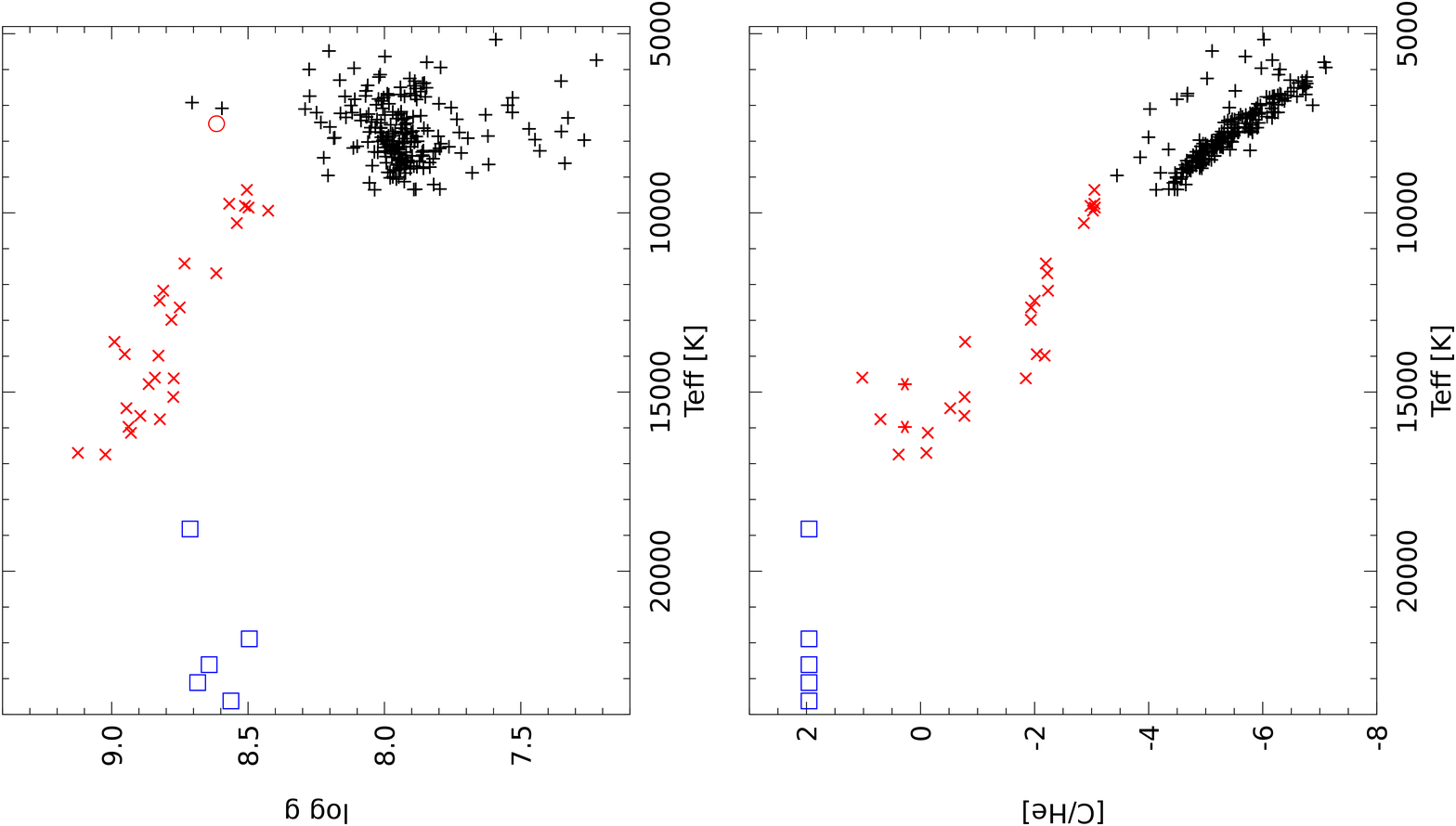}
\caption{Analysis results for the complete DQ sample.Top panel:
  \hbox{$\log g$}\ as a function of \hbox{$T\sb{\rm eff}$}\ for DQ
  (black crosses), wDQ (red xses), and hDQ (blue squares). The large
  red circle is the average result for peculiar DQs as explained in
  sect.~\ref{discussion}. 
  Bottom panel: [C/He] abundance ratios, with the same symbols as in the top
  panel, except for the two red asterisks, which are warm DQ with
  individually determined lower limit on [C/He]. The [C/He] for the
  hot DQ were held fixed and are typical lower limits.}
\label{figresult}
\end{figure}

\section{Results}
The objects showing Swan bands were fitted with the low temperature
grid, the others with the hot grid.

\subsection{The classical DQ}
Table~\ref{tabdq} presents the results for the atmospheric parameters
of the 221 DQ white dwarfs in the final sample. The range of
temperatures and abundances agrees with previous studies of large
samples like \citet{Dufour.Bergeron.ea05} and
\citet{Koester.Knist06}. With the exception of a figure in
\citet{Coutu.Dufour.ea18} this is, however, the first time that large
numbers of surface gravities (or masses) are presented for these
objects. There is one cool DQ identified in
\citet{Coutu.Dufour.ea18} which is also in our sample:
SDSSJ1355+3636, with \hbox{$T\sb{\rm eff}$}\ = 8040\,K, \hbox{$\log g$}\
= 7.91, [C/He] = $-5.04$. While the atmospheric parameters agree
quite well with our solution in Table~\ref{tabdq}, the [C/He] ratio we
obtain is $\approx 0.7$\,dex higher.

The gravities of the cool DQ are comparable to those of other cool
helium-dominated objects, as is demonstrated in
Table~\ref{tabcomparison}. The DZ, DB, and DC objects were drawn
  from the samples in \citet{dr7,dr10,dr12,dr14}, but the parameters
  were redetermined for this study. A significant fraction of DB white
  dwarfs is known to contain traces of hydrogen (spectral type DBA),
  and \citet{Bergeron.Dufour.ea19} have argued that using pure helium
  atmospheres for cool DC can lead to artificially higher masses. We
  have used different grids with [H/He] between $-3.0$ and $-6.0$ for
  the fitting. The results in the table were obtained with $-4.0$,
  which we have preferred because the resulting average masses are
  almost constant from 15000\,K down to 6000\,K. The DZ were fitted
  with [H/He]=$-4.5$, but in this case the effect of the hydrogen is
  not important, since the free electrons come from the metals.

The average masses between the three types of helium-rich
objects are slightly different, with the DQ having the smallest masses
and DZ and DC about 0.06 resp. 0.10\,dex larger. Considering the
extreme physical conditions in cool helium-rich atmospheres, the
uncertainties of the amount of hydrogen present, and the uncertain
reddening, we will not speculate on possible reasons for this
difference but refer further discussion to future work.

\subsection{The warm DQ (wDQ)}
These objects are characterized by \ion{C}{i} lines and occasionally
at the cool end with weak C$_2$ bands. Unfortunately there appear to
be severe problems with the atomic data of these lines. Our main
source was the Vienna Atomic Line Database
\citep[VALD,][]{Kupka.Piskunov.ea99, Kupka.Ryabchikova.ea00,
  Piskunov.Kupka.ea95}
\footnote{\url{http://vald.astro.univie.ac.at/~vald3/php/vald.php}}.
Spectral fits with theoretical models agree reasonably in the red part
of the spectrum $\lambda>4600$\,\AA. In the blue part, however, many
lines are predicted much stronger than they appear in the observed
spectrum. In the National Institute for Standards and Technology
\citep[NIST,][]{Kramida.Ralchenko.ea18} database\footnote{
  \url{https://www.nist.gov/pml/atomic-spectra-database}} are only two
lines in this range with $\log\ gf$ values published, none with class
A quality. The sources are \citet{Haris.Kramida17} and
\citet{Luo.Pradhan89}. In a literature search we found only very few
data, but some indication that the VALD data can be quite
wrong. \citet{Victor.Escalante88} give $\log\ gf$ values of $-2.143$
for the 4270.221\,\AA\ line, whereas the VALD value is $-1.637$, and
NIST has $-1.5$. For the 4892.019\,\AA\ line the same source has
$-3.602$, compared to $-1.541$ in VALD. We have replaced those and a
few other lines from the same source, and arbitrarily scaled the
values for several strong lines in the blue region to be compatible
with the fit to the red part. This can be considered as an empirical
$gf$ determination; the argument for this procedure is to make the
theoretical photometry more reliable. For the spectral fitting we used
only the region longward of 4600\,\AA.

As we determine \hbox{$T\sb{\rm eff}$}\ and \hbox{$\log g$}\ through
the photometry, the results do not depend much on the [C/He] ratio. On
the other hand, keeping \hbox{$T\sb{\rm eff}$}\ and \hbox{$\log
  g$}\ fixed when analyzing the spectra, the only remaining variable
is the abundance ratio. The carbon is diluted in the helium, weakening
the lines and leading to the abundance determination. As long as
[C/He] $\leq -1.0$ this works well. However, at higher ratios the
influence of helium weakens and the errors get much larger, up to
0.73\,dex. In two cases (SDSSJ1448+0519, SDSSJ1728+5558) this lead to
best fit models, which predicted strong He lines, that are not
observed. We have determined individual lower limits of [C/He] and
continued the iteration for \hbox{$T\sb{\rm eff}$}\ and
\hbox{$\log g$}\ with these limits fixed. These objects are indicated with
asterisks in Fig.~\ref{figresult}

The results of this analysis are shown in Table~\ref{tabwdq} and
Fig.~\ref{figresult}.  Apart from SDSSJ1448+0519 shown in
Fig.~\ref{Figcoutu}, there is one other warm DQ in our sample also
identified in \citet{Coutu.Dufour.ea18}, SDSSJ1215+4700 with
\hbox{$T\sb{\rm eff}$}\ = 12349\,K, \hbox{$\log g$}\ = 8.80, [C/He] =
$-2.27$. Our temperature is 1600\,K higher, the other parameters agree
reasonably well.

Many of the wDQ show Balmer lines of hydrogen. We therefore had
calculated the grid with trace hydrogen of [H/He] = $-4.5$, a
typical value for DBs. Some objects, however, show even much stronger
\hbox{H$\alpha$}\ than in the best fitting model, one of the strongest
being SDSSJ1622+3004. We determined the H abundance as [H/He ] = $-2.00$
and verified that this did not change the model spectrum except close
to the Balmer lines. The fit results for the wDQ are therefore not
affected. In a subsequent paper we will study the extent of the
convection zones beyond the bottom of the atmospheres (all models in
our grids are convective down to the bottom of the model) with
envelope models. We will then also determine accurate hydrogen
abundances to determine the total masses of H and He in the convection
zone. This could help to understand the origin and evolution of DQ
stars.

\subsection{Hot DQ (hDQ)}
We have arbitrarily identified the objects with \hbox{$T\sb{\rm eff}$}\
$>18000$\,K as hot DQs, since there is a continuous
transition in the strength of \ion{C}{ii} lines (e.g. 4621, 5146,
5891, 5893, 6153, 6463, 6579, 6584\,\AA) compared to \ion{C}{i}
lines. The complete absence of He lines in all hDQ translates in lower
limits of [C/He] = 1.0 to 2.0, depending on the S/N of the
spectrum. Since the fit results do not change with further increase of
this ratio we have used a fixed value of [C/He] = 2.0 for the fit.

\section{Discussion and conclusions\label{discussion}}
Detailed results of the analysis are presented in Tables~\ref{tabdq},
\ref{tabwdq}, and \ref{tabhdq}, but the main result is seen in
Fig.~\ref{figresult}. In the bottom panel, showing the abundances as
function of temperature, the classical DQs form a very well defined
dominant sequence from 9000\,K, [C/He]=$-4.5$ down to 5000\,K,
[C/He]=$-7.0$. This agrees with previous findings in the literature,
e.g. \citet{Pelletier.Fontaine.ea86,Dufour.Bergeron.ea05,Koester.Knist06}. A
number of objects are located in a range about 1.0-1.5\,dex in
abundance above this sequence, and there may be another sequence at
the upper limit of this range, although the evidence is not too
strong.  Going to higher \hbox{$T\sb{\rm eff}$}, the upper sequence
continues with the warm DQs up to 17000\,K, [C/He] around 0.0. This is
surprising since the warm DQs have been considered as helium-dominated
like the classical DQ \citep[e.g.][]{Coutu.Dufour.ea18}. Among the
hottest group is SDSSJ1448+0519 mentioned in the introduction and
shown in Fig.~\ref{Figcoutu}, which clearly cannot have as much helium
as found in \citet{Coutu.Dufour.ea18}. This group seems to be a
transition to the hot DQ, which are clearly carbon dominated, although
the helium content cannot be determined from the available spectra;
the squares in the figure are approximate lower limits for carbon.

Taking this bottom panel at face value, this would argue for a
continuous transition of very carbon-rich spectra to ever lower C
abundances with decreasing temperature. It is tempting to identify
this with some convection zones changing in depth with changes of
\hbox{$T\sb{\rm eff}$}\, but there are contradictions to our current
understanding of He layer convection, which shows a maximum depth
around 10000\,K \citep{Fontaine.Brassard05}.

Even more puzzling is the top panel of Fig.~\ref{figresult}. While the
cool DQ seem to have normal surface gravities around 8.0 and thus
normal masses, this is clearly not true for warm and hot DQ. The
current assumption was that the warm DQ have about the same masses as
the cool ones, but we find instead a sequence increasing from
\hbox{$\log g$}\ $\geq$ 8.5 to 9.1 with \hbox{$T\sb{\rm eff}$}\ from
10000 to 18000\,K, whereas the hot DQ have only \hbox{$\log
  g$}\ $\approx$ 8.5-8.7. The high \hbox{$\log g$}\ values for the
range 13000-17000\,K are the most puzzling result of our
study. Considering the problems with the atomic data for \ion{C}{i}
discussed above, one might question the validity of our analysis.  As
a test we have used the spectra also for the temperature determination
and the photometry only for the surface gravity, as in the hot
DQ. There are small differences in the parameters obtained, but the
general picture with the high \hbox{$\log g$}\ around 15000\,K
remains.

We note that the \hbox{$\log g$}\ determination via photometry and
parallax is very robust: assuming that the correct temperature for a
15000\,K determination were 13000\,K, i.e. an error of 2000\,K, the
surface gravity would only go down from 9.0 to 8.83, still
significantly higher than at higher and lower \hbox{$T\sb{\rm eff}$}.
One might argue that the mass-radius relation used -- for
He atmosphere white dwarfs -- is not appropriate, when [C/He]
$\approx$0.0. However, at 15000\,K, \hbox{$\log g$}\ = 9.0 the radius
changes very little between helium atmosphere
(0.00571\,\hbox{R$\sb{\odot}$}), hydrogen atmosphere
(0.00573\,\hbox{R$\sb{\odot}$}), or even the zero-temperature
Hamada-Salpeter radius (0.00570\,\hbox{R$\sb{\odot}$}\ for carbon,
\citet{Hamada.Salpeter61})

The problems with the \ion{C}{i} atomic data, some differences between
our results and \citet{Coutu.Dufour.ea18} and the unexpected surface
gravities for the warm DQ indicate that improvements to the model
calculations may still be necessary.  In spite of these remaining
uncertainties it is clear that there are two separate groups of carbon
white dwarfs. The classical DQ can be understood as dredge-up of the
helium convection zone from the tail of the C/He transition layer
\citep{Koester.Weidemann.ea82, Pelletier.Fontaine.ea86}; the decrease
of the carbon abundance with decreasing temperature is a result of the
depth of the convection zone decreasing again below the maximum near
10000\,K, as predicted by model calculations
\citep{Fontaine.Brassard05}.  If the high masses of warm and hot DQ
(which together may form the second group) are confirmed, they cannot
be the progenitors of the cool DQs. There are only two cool DQ with
\hbox{$\log g$}\ $>$ 8.50. This number seems too small because of the
slowing down of the evolution at low temperatures; a final conclusion,
however, would need a population synthesis taking into account the
accurate evolution including crystallization and Debye cooling, as
well as the observational detection biases in the SDSS.

Because our models do not adequately describe the spectra of the
peculiar DQ, we cannot determine reliable temperatures for them. The
presence of strong Swan bands indicates a range from 10000-5000\,K. We
have therefore used a fixed temperature of 7500\,K, [C/He] = $-5.00$ and
used the photometry to estimate the surface gravities. If the real
\hbox{$T\sb{\rm eff}$} is higher, the real \hbox{$\log g$} will be
higher than our result, and vice versa for lower real \hbox{$T\sb{\rm
    eff}$}. This indicates that the average result could be a
reasonable indicator of the typical gravities of the DQpec. This
average for the 21 DQpec in the sample is 8.625, with a distribution
width of 0.298. This value is shown as a large circle in
Fig.\ref{figresult}, confirming the possible identification of the
DQpec as descendants of the hot and warm classes. The position close
to the two cool DQ with higher gravity suggests that the two objects
may in fact belong to the DQpec class.

The high gravity also supports the idea that the unusual shape and
shift of the molecular Swan bands is indeed due to the high He
densities in these massive white dwarfs \citep{Kowalski10}.  We
do not yet know the C/He ratio for these stars, but a comparison of
band strengths with those in normal DQs indicates helium-dominated
atmospheres with slightly larger [C/He] values as in the DQ.

Another open question is how the hot DBs with small traces of carbon
visible only in the ultraviolet \citep{Fontaine.Brassard05,
  Koester.Provencal.ea14} are related to the two groups. Except
  for one object in the latter study accretion as origin of the carbon
  is considered highly unlikely, either because no other metals are
  observed, or because the inferred C/Si ratio in the accreted matter
  is extremely large. This is quite different from what is observed in
  accreting DAZ and DBZ in the same temperature range
  \citep{Koester.Gaensicke.ea14}, leaving dredge-up of the carbon as
  plausible explanation. Since they do not share the very high masses
  with the hot DQ they could be related to the classical cool DQ.

In a subsequent paper we will study the evolution of the convection
zones in mixed H/He/C envelopes and the total amount of hydrogen and
helium present in them. Hopefully that can shed some light on the
origin and evolution of these puzzling stars.

\begin{acknowledgement}
We thank Patrick Dufour and Simon Coutu for sharing with us a poster
presented at the EUROWD-21 workshop in Austin 2018. We have used
  their results in the belief that they are public. Only very recently
  we realized that this paper is not available in the public archive
  of the conference. We acknowledge that \citet{Coutu.Dufour.ea18} was
  the first paper to determine carbon abundances for DQ using {\it
    Gaia} data. Our current study was motivated by their results.
This work was financed in part by the Coordena\c{c}\~ao de
Aperfei\c{c}oamento de Pessoal de N\'{\i}vel Superior - Brasil (CAPES)
- Finance Code 001, Conselho Nacional de Desenvolvimento
Cient\'{\i}fico e Tecnol\'ogico - Brasil (CNPq), and Funda\c{c}\~ao de
Amparo \`a Pesquisa do Rio Grande do Sul (FAPERGS) - Brasil.  This
research has made use of NASA's Astrophysics Data System Bibliographic
Services, SIMBAD database, operated at CDS, Strasbourg, France, the
NIST and VALD atomic databases, and public data from the Sloan Digital
Sky Survey and the {\it Gaia} Mission.
\end{acknowledgement}  

%\bibliographystyle{aa} % style aa.bst
%\bibliography{wdnew} % your references Yourfile.bib

\begin{appendix}
\section{Full tables of objects and results}

\longtab[1]{
\begin{longtable}{ccrcccl}
\caption{\label{taballobjects} Selected sample of 304 stars with
  carbon features}\\
\hline\hline
  SDSS~J     &            p-m-f   &      $\pi$ [msec] &  g   &     G  &   S/N& Type \\
\hline
\endfirsthead
\caption{continued.}\\
\hline\hline
  SDSS~J     &            p-m-f   &      $\pi$ [msec] &  g   &     G  &   S/N& Type \\
\hline
\endhead
\hline
\endfoot
%%               % total 304  221 DQ      26 wDQ  7 hDQ   29 dC  21 DQpec
000011.66$-$085008.3 &  7167-56604-0752 &   4.825 &  19.111 &  19.056 &  17.8 &    DQ\\
000705.02$+$282104.2 &  2824-54452-0602 &   5.420 &  19.702 &  19.590 &  13.6 &    DQ\\
001523.91$+$030909.0 &  4298-55511-0906 &   5.020 &  20.028 &  19.933 &   8.5 &    DQ\\
001908.63$+$184706.0 &  7592-56947-0123 &   6.594 &  19.123 &  19.160 &   9.7 &    wDQ\\
002531.50$-$110800.9 &  0653-52145-0086 &   9.255 &  17.999 &  17.949 &  20.5 &    DQ\\
003328.58$+$041834.6 &  4303-55508-0562 &   7.497 &  18.720 &  18.648 &  26.2 &    DQ\\
004440.31$+$125923.3 &  6202-56266-0820 &   5.732 &  19.923 &  19.792 &  11.1 &    DQ\\
005505.41$+$085030.6 &  4544-55855-0064 &  11.351 &  18.689 &  18.392 &  16.2 &    DQ\\
010647.92$+$151327.6 &  5131-55835-0736 &   2.826 &  18.851 &  19.025 &  18.0 &    hDQ\\
010748.20$+$010240.1 &  2328-53728-0444 &  10.714 &  18.832 &  18.591 &  25.0 &    DQ\\
011639.84$+$234644.5 &  5693-56246-0401 &  11.183 &  17.769 &  17.731 &  34.5 &    DQ\\
012723.62$+$004630.0 &  0399-51817-0099 &   6.879 &  18.758 &  18.728 &  12.1 &    DQ\\
014324.13$+$115455.9 &  4662-55590-0818 &   4.797 &  20.523 &  20.340 &   6.6 &    DQ\\
015433.57$+$004047.2 &  0403-51871-0268 &   8.807 &  18.677 &  18.575 &  13.4 &    DQ\\
015441.74$+$140308.0 &  0430-51877-0558 &  16.213 &  17.865 &  17.666 &  22.1 &    DQ\\
020022.76$+$071459.7 &  4531-55563-0089 &  10.203 &  18.834 &  18.660 &  14.3 &    DQ\\
020252.80$+$004400.1 &  7837-56987-0256 &   9.522 &  20.203 &  19.604 &  12.2 &    dC\\
020534.24$+$215600.1 &  7636-56989-0171 &   9.522 &  19.902 &  19.604 &  11.7 &    DQ\\
022909.51$+$251024.5 &  2399-53764-0244 &   7.258 &  18.870 &  18.825 &  18.5 &    DQ\\
023633.74$+$250348.9 &  2399-53764-0059 &   5.696 &  18.824 &  18.945 &  16.5 &    wDQ \\
023945.04$+$002745.0 &  3745-55234-0924 &   4.714 &  19.733 &  19.631 &  23.6 &    DQ\\
024802.27$+$340802.4 &  2398-53768-0283 &  13.035 &  18.826 &  18.485 &  20.2 &    DQ\\
025357.12$+$341436.8 &  2398-53768-0154 &   3.915 &  19.781 &  19.674 &   9.0 &    DQ\\
032054.11$-$071625.4 &  0460-51924-0236 &   7.975 &  19.747 &  19.384 &   9.2 &    DQ\\
033004.02$+$051450.8 &  2339-53729-0036 &  10.226 &  19.647 &  19.087 &  13.5 &    dC\\
033218.23$+$003722.5 &  2069-53376-0148 &   9.040 &  18.389 &  18.333 &  24.9 &    DQ\\
035233.12$+$094641.1 &  2697-54389-0068 &   7.721 &  19.163 &  18.017 &  20.5 &    dC\\
041601.25$+$071308.9 &  2826-54389-0369 &   6.515 &  19.549 &  19.377 &  13.8 &    DQ\\
072339.12$+$390839.0 &  3657-55244-0476 &   8.476 &  17.795 &  17.864 &  31.0 &    DQ\\
073048.27$+$345729.8 &  2053-53446-0102 &   0.719 &  17.715 &  16.982 &  28.2 &    dC\\
073139.60$+$663618.1 &  2944-54523-0333 &   4.707 &  19.423 &  19.432 &  13.0 &    DQ\\
073703.83$+$645524.6 &  2944-54523-0244 &   5.097 &  19.619 &  19.525 &  11.1 &    DQ\\
074008.45$+$181047.2 &  2915-54497-0274 &   7.241 &  19.581 &  19.433 &  12.8 &    DQ\\
074204.75$+$434835.6 &  3669-55481-0722 &   7.164 &  18.815 &  18.752 &  21.1 &    DQ\\
074252.35$+$241140.5 &  4470-55587-0840 &   5.848 &  19.066 &  19.058 &  16.1 &    DQ\\
075000.55$+$232945.8 &  2916-54507-0474 &   5.347 &  20.167 &  20.084 &   9.4 &    DQ\\
075059.15$+$132855.4 &  4502-55569-0218 &   6.156 &  19.017 &  18.973 &  18.3 &    DQ\\
075200.65$+$301943.4 &  3752-55236-0652 &   5.647 &  20.171 &  20.103 &   9.0 &    DQ\\
075230.82$+$444749.9 &  3671-55483-0028 &  14.087 &  16.891 &  16.893 &  51.2 &    DQ\\
080236.30$+$414741.8 &  3683-55178-0162 &   4.257 &  20.250 &  20.169 &   9.2 &    DQ\\
080405.76$+$075055.1 &  2076-53442-0408 &   4.668 &  19.711 &  19.649 &  11.2 &    DQ\\
080455.42$+$171443.5 &  2081-53357-0171 &  14.947 &  19.079 &  18.521 &  12.3 &    DQpec\\
080558.84$+$072448.5 &  2076-53442-0505 &  12.799 &  19.565 &  18.984 &  13.5 &    DQpec\\
080708.49$+$194950.0 &  4481-55630-0323 &   5.663 &  18.844 &  19.032 &  21.0 &    wDQ\\
080843.15$+$464028.7 &  0438-51884-0063 &   9.668 &  20.399 &  19.558 &   6.3 &    DQ\\
081323.32$+$304744.1 &  0861-52318-0053 &   7.248 &  18.912 &  18.881 &  12.6 &    DQ\\
081448.31$+$245540.3 &  4463-55868-0140 &  16.968 &  16.921 &  16.877 &  14.1 &    DQ\\
081839.24$+$010227.5 &  2077-53846-0575 &   3.609 &  17.598 &  17.869 &  23.6 &    hDQ\\
082219.23$+$202325.8 &  1927-53321-0125 &  23.299 &  16.274 &  16.129 &  49.3 &    DQ\\
082626.78$+$470911.3 &  7325-56717-0705 &   1.221 &  19.232 &  17.838 &  19.9 &    dC\\
082710.45$+$210053.0 &  4483-55587-0850 &   7.461 &  19.284 &  19.127 &  13.8 &    DQ\\
083310.19$+$363846.9 &  0864-52320-0314 &   7.038 &  19.038 &  18.923 &  11.0 &    DQ\\
083608.56$+$043757.1 &  3809-55533-0504 &   5.786 &  18.842 &  18.864 &  19.9 &    DQ\\
083618.13$+$243254.6 &  2330-53738-0627 &  10.226 &  19.524 &  19.087 &  15.5 &    DQpec\\
083637.79$+$481752.5 &  0550-51959-0433 &   8.254 &  18.809 &  18.683 &  15.8 &    DQ\\
083728.83$+$032129.0 &  3809-55533-0466 &   5.787 &  19.165 &  19.131 &  19.6 &    DQ\\
083815.75$+$112118.4 &  5284-55866-0786 &   7.045 &  19.263 &  19.148 &  15.6 &    DQ\\
083921.07$+$084250.3 &  5284-55866-0206 &   6.748 &  20.236 &  19.945 &   6.7 &    DQ\\
084001.22$+$452933.4 &  5163-55889-0646 &   9.516 &  18.298 &  18.251 &  24.3 &    DQ\\
084131.55$+$332915.5 &  0933-52642-0016 &  18.804 &  18.457 &  18.370 &  14.1 &    DQ\\
084531.38$+$614336.4 &  1875-54453-0184 &  13.082 &  17.499 &  17.458 &  29.9 &    DQ\\
084624.72$+$102406.0 &  2671-54141-0310 &   7.721 &  17.975 &  18.017 &  32.3 &    DQ\\
085030.31$+$070937.0 &  5289-55893-0088 &   4.161 &  19.341 &  19.393 &  17.6 &    DQ\\
085506.63$+$063904.7 &  1189-52668-0535 &  13.082 &  17.862 &  17.740 &  21.3 &    DQ\\
085626.94$+$451337.0 &  7327-56715-0052 &   4.804 &  19.965 &  20.045 &   8.3 &   wDQ\\
085709.01$+$060357.4 &  1189-52668-0027 &   8.116 &  18.450 &  18.402 &  13.7 &    DQ\\
085914.63$+$325712.2 &  1272-52989-0309 &  43.363 &  15.086 &  15.089 &  69.3 &    wDQ\\
090148.21$+$365402.6 &  8860-57458-0148 &   6.761 &  19.000 &  20.454 &  21.7 &    dC\\
090200.36$+$503723.1 &  0551-51993-0612 &   6.086 &  19.240 &  19.177 &  10.2 &    DQ\\
090449.73$+$395416.5 &  1199-52703-0595 &   6.395 &  19.560 &  19.449 &   7.8 &    DQ\\
090514.78$+$090426.5 &  5299-55927-0646 &  16.790 &  16.732 &  16.694 &  48.3 &    DQ\\
090632.17$+$470235.8 &  0898-52606-0565 &  11.654 &  20.339 &  19.505 &   5.1 &    DQ\\
091406.56$+$063254.2 &  1193-52652-0528 &   1.262 &  19.504 &  17.986 &  13.0 &    dC\\
091500.58$+$201903.1 &  5768-56017-0666 &   5.167 &  19.158 &  19.224 &  16.9 &    DQ\\
091602.83$+$101109.7 &  5302-55896-0023 &  25.769 &  15.806 &  15.765 &  53.3 &    DQ\\
091922.26$+$023604.4 &  3822-55544-0966 &   6.399 &  18.980 &  19.077 &  19.6 &    wDQ\\
092011.52$+$530616.9 &  7289-57039-0974 &   4.407 &  19.924 &  19.875 &   9.2 &    DQ\\
092047.01$+$360352.8 &  4644-55922-0898 &   6.487 &  19.148 &  19.084 &  13.6 &    DQ\\
092153.48$+$342136.5 &  5810-56358-0620 &   5.512 &  18.599 &  18.583 &  23.3 &    DQ\\
092613.46$+$472521.1 &  0900-52637-0041 &  10.701 &  18.390 &  18.267 &  19.6 &    DQ\\
092909.03$+$331011.7 &  1593-52991-0094 &  11.009 &  18.802 &  18.437 &  12.8 &    DQ\\
093017.14$+$295940.7 &  2914-54533-0304 &   6.348 &  18.792 &  18.761 &  25.1 &    DQ\\
093448.86$+$115854.3 &  5313-55973-0175 &   6.263 &  18.711 &  18.710 &  24.2 &    DQ\\
093537.57$+$241708.7 &  2294-53733-0044 &   7.897 &  18.215 &  18.251 &  28.1 &    DQ\\
093638.04$+$060709.6 &  4871-55928-0024 &   6.171 &  19.135 &  19.273 &  12.8 &    wDQ\\
093937.62$+$520142.8 &  5724-56364-0374 &  13.593 &  17.162 &  17.171 &  37.7 &    DQ\\
094004.64$+$021022.6 &  0477-52026-0493 &  19.232 &  17.262 &  17.129 &  33.0 &    DQ\\
094058.34$+$384350.2 &  3224-54849-0517 &   6.891 &  18.752 &  18.660 &  18.5 &    DQ\\
094115.18$+$090154.4 &  1304-52993-0009 &  11.174 &  16.763 &  16.791 &  37.9 &    DQ\\
094138.09$+$441457.7 &  4695-55957-0208 &   7.433 &  18.670 &  18.632 &  18.4 &    DQ\\
094415.10$+$582749.3 &  5715-56657-0715 &   6.433 &  18.584 &  18.628 &  21.8 &    DQ\\
094511.56$+$503214.4 &  7292-56709-0649 &   4.407 &  20.216 &  19.875 &  13.0 &    dC\\
094857.89$+$123243.0 &  1742-53053-0616 &  12.428 &  18.300 &  18.147 &  17.8 &    DQ\\
095017.10$+$531515.0 &  0769-52282-0135 &  36.308 &  15.176 &  15.131 &  43.5 &    DQ\\
095137.61$+$624348.7 &  2403-53795-0577 &   5.920 &  18.855 &  18.833 &  24.3 &    DQ\\
095545.86$+$443640.1 &  8286-57062-0866 &   4.034 &  16.985 &  20.247 &  45.6 &    dC\\
095837.00$+$585303.0 &  5719-56014-0114 &   6.014 &  18.758 &  18.958 &  12.9 &    wDQ\\
100523.61$-$011430.7 &  3769-55240-0546 &   5.089 &  19.307 &  19.305 &  15.2 &    DQ\\
100531.23$+$053809.2 &  0995-52731-0069 &   1.862 &  19.007 &  17.282 &  19.3 &    dC\\
100958.70$+$010313.2 &  0502-51957-0216 &   0.762 &  18.138 &  17.201 &  29.5 &    dC\\
101009.56$+$230016.8 &  6458-56274-0618 &   7.284 &  18.422 &  18.468 &  22.8 &    DQ\\
101219.90$+$004019.7 &  0502-51957-0095 &   8.386 &  17.710 &  17.762 &  28.4 &    DQ\\
101336.56$+$435054.8 &  4561-55614-0788 &   6.939 &  18.625 &  18.618 &  16.8 &    DQ\\
101509.60$+$351813.4 &  4568-55600-0256 &   5.781 &  19.001 &  18.978 &  16.3 &    DQ\\
101750.38$+$373637.5 &  1427-52996-0216 &   8.065 &  18.913 &  18.853 &  12.1 &    DQ\\
101800.00$+$083820.3 &  1237-52762-0621 &  22.825 &  16.410 &  16.347 &  36.6 &    DQ\\
102211.67$+$284507.3 &  2351-53772-0170 &   7.158 &  19.127 &  19.104 &  12.1 &    DQ\\
102635.06$+$272507.3 &  6463-56340-0894 &   6.625 &  18.858 &  18.832 &  17.9 &    DQ\\
102705.08$+$121836.1 &  1747-53075-0241 &  12.428 &  18.613 &  18.147 &  13.7 &    DQ\\
102801.75$+$351258.0 &  1958-53385-0558 &  13.533 &  19.389 &  18.739 &  10.6 &    DQ\\
102836.21$+$253732.8 &  2349-53734-0535 &   9.112 &  19.008 &  18.783 &  12.0 &    DQ\\
103102.52$+$221714.7 &  5873-56035-0828 &   7.359 &  20.192 &  20.115 &   7.3 &    DQ\\
103205.85$+$210131.2 &  5873-56035-0852 &   5.993 &  19.124 &  19.146 &  15.0 &    DQ\\
103210.68$+$451930.0 &  4691-55651-0758 &   7.931 &  18.080 &  18.114 &  30.0 &    DQ\\
104031.99$+$182206.6 &  5886-56034-0578 &   5.079 &  20.312 &  20.164 &   7.0 &    DQ\\
104204.26$+$583348.1 &  7095-56625-0462 &   7.947 &  19.084 &  18.983 &  10.5 &    DQ\\
104505.66$+$213447.2 &  5874-56039-0954 &   6.497 &  19.114 &  19.043 &  12.7 &    DQ\\
104906.61$+$165923.7 &  5352-56269-0732 &   5.759 &  19.298 &  19.423 &  20.4 &    wDQ\\
105119.71$+$301946.7 &  1981-53463-0520 &   1.924 &  20.085 &  18.428 &   9.1 &    dC\\
105248.66$+$591150.6 &  7095-56625-0848 &   8.786 &  18.320 &  18.275 &  37.8 &    DQ\\
105817.66$+$284609.3 &  2870-54534-0429 &   6.480 &  19.334 &  19.346 &  13.1 &   wDQ\\
105854.34$+$344018.1 &  4631-55617-0148 &   6.035 &  20.102 &  19.969 &  12.1 &    DQ\\
110058.04$+$175807.0 &  2485-54176-0119 &   7.060 &  18.589 &  18.732 &  13.5 &    wDQ\\
110140.41$+$521806.7 &  6706-56385-0856 &   5.219 &  19.220 &  19.204 &  14.3 &    DQ\\
110406.69$+$203528.6 &  6428-56279-0282 &   5.504 &  17.212 &  17.496 &  40.4 &    hDQ\\
110831.50$+$134950.5 &  5360-55983-0034 &  11.130 &  18.123 &  18.041 &  25.7 &    DQ\\
111336.00$+$445505.1 &  6649-56364-0050 &   6.681 &  18.898 &  18.865 &  17.4 &    DQ\\
111341.33$+$014641.7 &  0510-52381-0184 &  22.947 &  19.205 &  18.527 &  10.9 &    DQpec\\
112023.61$+$425200.7 &  1440-53084-0408 &   9.139 &  18.616 &  18.536 &  12.0 &    DQ\\
112324.94$+$352341.5 &  4618-55600-0334 &   5.542 &  19.986 &  19.791 &   6.9 &    DQ\\
112329.64$+$620714.4 &  3328-54964-0497 &   5.959 &  19.495 &  19.327 &  13.6 &    DQ\\
112348.02$+$334713.8 &  4619-55599-0440 &   7.728 &  19.044 &  18.983 &  13.9 &    DQ\\
112604.28$+$441938.6 &  3215-54861-0257 &   8.237 &  19.093 &  18.978 &  16.0 &    DQ\\
112702.85$+$190912.8 &  2498-54169-0166 &   1.249 &  19.574 &  18.190 &  11.9 &    dC\\
112731.86$-$021242.2 &  3233-54891-0206 &   6.891 &  17.633 &  18.660 &  41.9 &    dC\\
112804.62$-$072413.5 &  2876-54581-0472 &   2.924 &  18.867 &  19.743 &  35.2 &    dC\\
113104.23$+$184558.8 &  5879-56047-0298 &   6.496 &  18.804 &  18.762 &  23.2 &    DQ\\
113300.58$+$190056.5 &  5879-56047-0178 &   5.432 &  19.127 &  19.096 &  19.2 &    DQ\\
113534.62$+$572451.7 &  1310-53033-0485 &  17.103 &  16.480 &  16.371 &  40.3 &    DQ\\
114006.29$+$182401.9 &  5891-56034-0946 &  10.553 &  17.954 &  17.987 &  23.7 &    wDQ\\
114056.78$+$154014.3 &  1755-53386-0607 &   8.632 &  19.275 &  19.021 &   9.9 &    DQ\\
114059.88$+$073529.9 &  5377-55957-0240 &   6.214 &  19.166 &  19.270 &  14.9 &    wDQ\\
114136.53$+$383611.9 &  1997-53442-0386 &  12.641 &  19.920 &  19.107 &   7.2 &    DQpec\\
114256.66$+$035208.0 &  4766-55677-0262 &   5.481 &  19.261 &  19.169 &  15.2 &    DQ\\
114851.68$-$012612.8 &  0329-52056-0578 &  14.681 &  17.403 &  17.406 &  21.1 &    wDQ\\
115149.82$+$452730.3 &  6642-56396-0596 &   9.401 &  17.807 &  17.833 &  24.0 &    DQ\\
115933.10$+$130031.6 &  3285-54948-0513 &  16.104 &  18.137 &  17.818 &  29.5 &    DQpec \\
120027.73$+$225212.9 &  2643-54208-0469 &   2.320 &  18.976 &  19.258 &  10.8 &    hDQ\\
120049.38$+$423726.1 &  1447-53120-0621 &  10.826 &  19.270 &  18.990 &   8.8 &    DQ\\
120154.70$+$340055.8 &  2089-53498-0318 &  24.677 &  17.754 &  17.343 &  33.9 &    DQ\\
120331.77$+$645059.6 &  6975-56720-0446 &  11.458 &  17.528 &  17.656 &  38.5 &    wDQ\\
121149.26$+$044050.8 &  4749-55633-0722 &   6.493 &  19.129 &  19.058 &  15.6 &    DQ\\
121211.13$+$545221.1 &  6688-56412-0854 &  10.581 &  17.869 &  17.840 &  28.5 &    DQ\\
121510.66$+$470010.4 &  6640-56385-0532 &   6.754 &  18.764 &  18.962 &  13.6 &   wDQ\\
122257.72$+$604100.9 &  6968-56443-0588 &   5.198 &  19.187 &  19.194 &  15.6 &    DQ\\
122545.88$+$470613.0 &  1451-53117-0035 &   8.880 &  19.576 &  19.147 &   6.6 &    DQ\\
122854.55$+$480620.6 &  6671-56388-0004 &   5.839 &  19.624 &  19.480 &   7.1 &    DQ\\
123313.48$+$082403.1 &  1627-53473-0578 &  12.248 &  18.666 &  18.384 &  14.1 &    DQpec\\
123534.88$+$391820.0 &  3970-55591-0354 &  12.562 &  17.113 &  17.182 &  35.0 &    DQ\\
123752.12$+$415625.8 &  1454-53090-0146 &  27.659 &  17.757 &  17.150 &  24.7 &    DQpec\\
124034.89$-$014459.7 &  2922-54612-0392 &   4.767 &  19.363 &  19.352 &  17.9 &    DQ\\
124338.97$+$360729.0 &  2022-53827-0238 &   8.559 &  18.190 &  18.173 &  16.2 &    DQ\\
124433.81$+$270923.9 &  2238-54205-0229 &   6.139 &  18.340 &  18.270 &  16.9 &    DQ\\
124701.79$+$411340.8 &  4703-55617-0438 &   8.387 &  18.249 &  18.246 &  26.2 &    DQ\\
124733.70$+$491524.8 &  2923-54563-0222 &   9.114 &  19.400 &  19.094 &  14.6 &    DQ\\
124739.05$+$064604.6 &  1790-53876-0110 &  19.058 &  20.041 &  18.854 &   4.0 &    DQpec\\
124911.75$+$340706.0 &  3971-55322-0018 &  13.949 &  17.087 &  17.111 &  39.6 &    DQ\\
125017.90$+$252427.6 &  2661-54505-0444 &   3.537 &  18.199 &  16.501 &  27.3 &    dC\\
125151.56$+$464627.2 &  1457-53116-0215 &   9.373 &  17.818 &  17.841 &  20.0 &    DQ\\
125245.44$+$194311.2 &  2615-54483-0381 &  11.895 &  17.757 &  17.713 &  26.6 &    DQ\\
125359.61$+$013925.6 &  0523-52026-0252 &   5.624 &  19.074 &  19.062 &   9.7 &    DQ\\
125829.51$+$445206.9 &  1373-53063-0638 &   2.564 &  19.986 &  18.106 &   8.9 &    dC\\
130201.30$+$092351.7 &  3234-54885-0143 &   8.800 &  18.702 &  18.568 &  18.6 &    DQ\\
130945.51$+$444540.9 &  6621-56366-0844 &   9.137 &  18.970 &  18.836 &  17.4 &    DQpec\\
131534.80$+$471107.1 &  6625-56386-0556 &  11.834 &  17.760 &  17.690 &  29.7 &    DQ\\
131640.71$+$081058.7 &  5423-55958-0996 &  11.401 &  18.572 &  18.392 &  19.6 &    DQ\\
131930.71$+$140136.4 &  5425-56003-0148 &   6.901 &  18.967 &  18.891 &  18.3 &    DQ\\
131953.49$+$084422.0 &  5429-55979-0598 &   9.212 &  18.060 &  18.032 &  28.0 &    DQ\\
132232.05$+$373032.9 &  2102-54115-0356 &   7.212 &  18.985 &  18.932 &  15.8 &    DQ\\
132825.28$+$364016.5 &  3983-55603-0800 &   7.931 &  18.450 &  18.443 &  30.2 &    DQ\\
132931.08$+$074651.9 &  1801-54156-0324 &  14.252 &  17.223 &  17.265 &  31.9 &    DQ\\
133127.01$+$670420.0 &  6825-56717-0629 &   5.790 &  18.888 &  18.898 &  18.1 &    DQ\\
133151.38$+$372754.8 &  3984-55333-0118 &   7.466 &  18.264 &  18.460 &  25.6 &    wDQ\\
133221.56$+$235502.2 &  5995-56093-0918 &   4.409 &  19.157 &  19.366 &  11.4 &    wDQ\\
133313.73$+$235722.0 &  5995-56093-0958 &   7.596 &  18.756 &  18.736 &  14.6 &    DQ\\
133359.86$+$001654.8 &  0298-51955-0492 &  24.453 &  19.420 &  18.533 &  11.3 &    DQpec\\
133420.09$+$162235.8 &  5434-56033-0075 &   8.448 &  18.135 &  18.142 &  24.7 &    DQ\\
133641.85$+$192056.2 &  5862-56045-0694 &   7.886 &  19.780 &  19.537 &   9.8 &    DQ\\
133901.47$+$501434.4 &  6744-56399-0626 &   7.592 &  18.337 &  18.350 &  20.6 &    DQ\\
133940.50$+$503613.5 &  6744-56399-0664 &   5.427 &  19.114 &  19.243 &  11.9 &    wDQ\\
134124.28$+$034628.7 &  4786-55651-0184 &   4.641 &  19.329 &  19.473 &  13.3 &    wDQ\\
134426.47$+$184930.9 &  2930-54589-0391 &   8.672 &  18.071 &  18.062 &  40.2 &    DQ\\
134707.57$+$381749.7 &  3852-55243-0630 &   6.398 &  19.174 &  19.160 &  19.5 &    DQ\\
134716.09$+$501942.1 &  6740-56401-0202 &   6.487 &  18.906 &  18.883 &  14.9 &    DQ\\
134747.94$+$152851.0 &  5441-56017-0848 &   6.789 &  19.133 &  19.113 &  15.8 &    DQ\\
135057.54$+$050841.0 &  0855-52375-0474 &   2.243 &  17.434 &  16.064 &  32.2 &    dC\\
135134.35$+$662315.4 &  6986-56717-0616 &   9.978 &  17.703 &  17.712 &  31.5 &    DQ\\
135201.98$+$221811.4 &  5869-56064-0716 &   5.898 &  19.250 &  19.188 &  13.1 &    DQ\\
135258.69$+$265853.7 &  6005-56090-0670 &   6.494 &  19.123 &  19.060 &  12.3 &    DQ\\
135409.79$+$121732.2 &  5443-56010-0064 &   4.631 &  19.486 &  19.416 &  12.2 &    DQ\\
135516.57$+$363613.2 &  3852-55243-0048 &  11.665 &  17.684 &  17.644 &  37.0 &    DQ\\
135628.25$+$000941.2 &  0301-51942-0231 &  12.797 &  18.602 &  18.346 &  15.2 &    DQ\\
135739.52$+$294921.5 &  2122-54178-0346 &   8.398 &  18.641 &  18.556 &  15.1 &    DQ\\
135810.42$+$055237.6 &  4863-55688-0462 &   5.863 &  19.047 &  18.990 &  14.8 &    DQ\\
140625.70$+$020447.0 &  0532-51993-0475 &   5.648 &  18.773 &  18.670 &  12.4 &    DQ\\
140632.42$+$014838.3 &  0532-51993-0234 &   8.856 &  18.427 &  18.358 &  14.0 &    DQ\\
140723.00$+$203918.4 &  5894-56039-0782 &   6.334 &  18.653 &  18.636 &  17.0 &    DQ\\
141648.86$+$301654.0 &  6497-56329-0888 &   4.456 &  20.341 &  20.293 &  11.7 &    DQ\\
141745.42$+$241222.6 &  2128-53800-0153 &   8.342 &  18.645 &  18.596 &  15.4 &    DQ\\
142032.63$+$531624.5 &  7028-56449-0881 &   6.135 &  19.371 &  19.334 &  11.0 &    DQ\\
142413.37$+$083316.5 &  5462-55978-0432 &   8.866 &  17.710 &  17.751 &  29.2 &    DQ\\
142531.97$+$180115.7 &  5897-56042-0259 &   7.575 &  18.600 &  18.565 &  18.8 &    DQ\\
142625.72$+$575218.6 &  6803-56402-0916 &   3.078 &  19.148 &  19.394 &  12.1 &    hDQ\\
142728.30$+$611026.4 &  0607-52368-0379 &  23.283 &  17.154 &  16.924 &  35.0 &    DQ\\
142812.00$+$443439.8 &  2932-54595-0125 &   8.672 &  19.001 &  18.062 &  22.7 &    dC\\
143144.83$+$375011.9 &  1381-53089-0599 &   6.321 &  19.337 &  19.118 &  10.8 &    DQ\\
143227.07$+$475110.7 &  6752-56366-0851 &   8.269 &  18.200 &  18.175 &  22.2 &    DQ\\
143437.81$+$225859.2 &  6016-56073-0466 &   4.836 &  18.972 &  19.192 &  16.6 &    wDQ\\
143534.01$+$531815.1 &  1327-52781-0413 &   5.005 &  19.021 &  19.212 &   8.4 &    wDQ\\
143550.13$+$212014.5 &  5900-56043-0959 &   4.999 &  19.302 &  19.284 &  11.4 &    DQ\\
144221.53$+$420250.2 &  6061-56076-0860 &   5.269 &  19.092 &  19.144 &  16.6 &    DQ\\
144537.07$+$185120.4 &  5903-56064-0342 &   5.526 &  19.103 &  19.082 &  15.4 &    DQ\\
144808.07$+$004755.9 &  0308-51662-0145 &  11.170 &  18.438 &  18.293 &  18.0 &    DQ\\
144854.80$+$051903.5 &  4858-55686-0082 &   8.316 &  18.007 &  18.190 &  26.3 &    wDQ\\
144954.84$+$291633.7 &  3878-55361-0654 &   6.270 &  19.215 &  19.145 &  14.4 &    DQ\\
145005.60$+$102833.8 &  3388-54947-0393 &   5.959 &  17.397 &  19.327 &  42.9 &    dC\\
145023.96$+$635053.8 &  2947-54533-0605 &   4.340 &  19.664 &  19.637 &   8.7 &    DQ\\
145725.27$+$210747.4 &  2150-54510-0288 &  11.992 &  19.148 &  18.741 &  12.8 &    DQpec\\
150126.78$+$210056.9 &  2150-54510-0135 &  10.843 &  18.695 &  18.239 &  16.6 &    DQ\\
150136.90$+$062720.6 &  1815-53884-0095 &   9.279 &  18.968 &  18.848 &  11.3 &    DQ\\
150245.71$-$022454.8 &  0922-52426-0221 &   1.145 &  19.187 &  17.972 &  12.2 &    dC\\
150901.11$+$162012.3 &  5484-56039-0232 &   6.238 &  18.850 &  18.809 &  17.8 &    DQ\\
151118.19$+$500800.7 &  1330-52822-0417 &   6.815 &  18.986 &  18.954 &   9.4 &    DQ\\
151417.00$+$142354.1 &  2766-54242-0126 &   8.679 &  19.714 &  19.343 &   6.7 &    DQ\\
151734.38$+$225619.1 &  3955-55678-0592 &  12.614 &  17.850 &  17.773 &  33.1 &    DQ\\
151942.63$+$385658.8 &  4979-56045-0654 &   5.918 &  19.045 &  19.027 &  16.1 &    DQ\\
152030.30$+$270321.5 &  3851-55302-0214 &   5.778 &  19.224 &  19.120 &  19.5 &    DQ\\
152702.59$+$275213.2 &  3959-55679-0676 &   7.521 &  19.271 &  19.180 &  14.9 &    DQpec\\
152812.03$+$513444.1 &  6721-56398-0969 &   9.630 &  18.429 &  18.359 &  20.1 &    DQ\\
152840.05$+$400335.0 &  2936-54626-0548 &   7.174 &  19.042 &  19.017 &  19.6 &    DQ\\
152854.10$+$044226.6 &  1835-54563-0179 &   7.043 &  19.122 &  19.012 &   8.2 &    DQ\\
153303.83$+$395818.3 &  1679-53149-0135 &   3.342 &  18.928 &  19.248 &  16.8 &    dC\\
153447.54$+$414559.4 &  1679-53149-0616 &  15.188 &  17.244 &  17.167 &  32.3 &    DQ\\
153732.19$+$004343.1 &  0315-51663-0422 &   2.618 &  19.471 &  17.600 &  15.1 &    dC\\
153746.05$+$133732.9 &  4890-55741-0439 &   8.507 &  18.385 &  18.323 &  22.5 &    DQ\\
154156.85$+$514421.3 &  0796-52401-0423 &   1.140 &  18.486 &  17.537 &  18.5 &    dC\\
154234.73$+$254438.7 &  3947-55332-0760 &   8.234 &  19.325 &  19.133 &  11.7 &    DQ\\
154238.94$+$131455.8 &  4890-55741-0172 &   8.556 &  18.450 &  18.396 &  26.9 &    DQ\\
154810.66$+$562647.7 &  0617-52072-0551 &   6.854 &  19.028 &  19.000 &  10.1 &    DQ\\
155155.26$+$082431.8 &  1727-53859-0442 &  13.901 &  18.018 &  17.872 &  20.2 &    DQ\\
155202.39$+$114829.1 &  2520-54584-0515 &   7.142 &  19.002 &  18.911 &  15.0 &    DQ\\
155206.42$+$391017.6 &  5192-56066-0472 &   8.797 &  18.696 &  18.603 &  13.6 &    DQ\\
155413.53$+$033634.5 &  0595-52023-0373 &   9.761 &  18.914 &  18.686 &  11.5 &    DQ\\
155432.04$+$113616.0 &  2520-54584-0630 &   5.714 &  19.060 &  19.023 &  13.6 &    DQ\\
160935.76$+$065509.9 &  1730-53498-0217 &  11.686 &  17.737 &  17.678 &  23.7 &    DQ\\
161018.90$+$303619.1 &  5009-55707-0838 &   6.268 &  19.113 &  19.059 &  16.7 &    DQ\\
161140.18$+$045127.0 &  2189-54624-0562 &  14.823 &  18.703 &  18.303 &  23.0 &    DQpec\\
161315.43$+$511608.2 &  6316-56483-0713 &   6.201 &  19.216 &  19.165 &  15.5 &    DQ\\
161414.12$+$172900.5 &  2177-54557-0460 &  18.426 &  18.671 &  18.012 &  21.3 &    DQpec\\
161653.36$+$392444.4 &  1336-52759-0572 &  10.687 &  18.325 &  18.203 &  17.1 &    DQ\\
161847.38$+$061155.2 &  1732-53501-0218 &  12.938 &  18.248 &  18.255 &  18.4 &    DQpec\\
161853.78$+$504800.8 &  3288-54908-0467 &  16.104 &  17.311 &  17.818 &  43.6 &    dC\\
161945.61$+$192841.8 &  4060-55359-0568 &   9.919 &  19.425 &  18.035 &  17.7 &    dC\\
162004.02$+$180912.3 &  4060-55359-0444 &   9.919 &  18.089 &  18.035 &  28.7 &    DQ\\
162153.79$+$225309.4 &  4184-55450-0463 &   8.110 &  18.521 &  18.473 &  21.1 &    DQ\\
162205.12$+$184956.7 &  4060-55359-0346 &   5.434 &  19.202 &  19.383 &  15.3 &    wDQ\\
162236.13$+$300454.5 &  4953-55749-0483 &  13.366 &  16.895 &  17.095 &  38.5 &    wDQ\\
162635.58$+$154441.6 &  4056-55357-0734 &   7.976 &  19.661 &  19.440 &  12.9 &    DQpec\\
162721.61$+$304320.2 &  4953-55749-0732 &   6.412 &  18.741 &  18.742 &  18.6 &    DQ\\
163546.48$+$284335.5 &  5201-55832-0412 &   7.425 &  19.147 &  19.062 &  13.2 &    DQ\\
163857.96$+$124420.6 &  4075-55352-0802 &   9.860 &  17.788 &  17.796 &  45.0 &    DQ\\
164143.26$+$483301.8 &  6318-56186-0724 &  10.937 &  18.096 &  18.036 &  27.9 &    DQ\\
165436.86$+$315754.4 &  1176-52791-0238 &  15.781 &  17.504 &  17.408 &  23.3 &    DQ\\
165538.46$+$372247.9 &  5198-55823-0844 &   6.789 &  18.491 &  18.521 &  20.7 &    DQ\\
171221.94$+$340609.9 &  4994-55739-0426 &   5.539 &  19.298 &  19.271 &  13.9 &    DQ\\
171341.84$+$324007.5 &  4998-55722-0536 &  14.362 &  17.360 &  17.283 &  36.8 &    DQ\\
172856.22$+$555822.8 &  0358-51818-0296 &  21.188 &  16.024 &  16.169 &  48.5 &    wDQ\\
180302.57$+$232043.3 &  2195-54234-0272 &  15.335 &  20.478 &  19.193 &   9.9 &    DQpec\\
183500.21$+$642916.9 &  2552-54632-0496 &  18.726 &  17.594 &  17.278 &  38.4 &    DQpec\\
204624.45$-$071519.1 &  0635-52145-0228 &   6.145 &  18.916 &  18.833 &  11.0 &    DQ\\
205316.34$-$070204.3 &  0636-52176-0267 &  10.190 &  19.210 &  18.818 &  10.9 &    DQ\\
213503.31$+$000318.4 &  0989-52468-0198 &   8.283 &  19.399 &  19.146 &   8.5 &    DQ\\
214058.23$+$004529.4 &  4196-55478-0444 &   6.479 &  19.369 &  19.265 &  17.8 &    DQ\\
214247.87$+$090858.8 &  4093-55475-0084 &   7.631 &  19.333 &  19.210 &  15.5 &    DQ\\
214944.08$+$203921.6 &  5949-56096-0402 &   4.727 &  19.094 &  19.083 &  16.9 &    DQ\\
215027.79$-$011351.0 &  4197-55479-0362 &  11.063 &  17.465 &  17.478 &  45.6 &    DQ\\
215652.94$+$055925.1 &  4096-55501-0804 &  10.286 &  18.680 &  18.556 &  15.4 &    DQ\\
215759.11$+$113729.1 &  5063-55831-0387 &   6.581 &  18.590 &  18.585 &  23.3 &    DQ\\
220029.09$-$074121.6 &  0717-52468-0462 &   5.029 &  17.781 &  17.975 &  21.6 &    hDQ\\
220746.17$+$013509.0 &  4316-55505-0170 &   8.647 &  19.467 &  19.276 &  12.0 &    DQ\\
221245.22$+$061323.4 &  2323-54380-0113 &   8.882 &  19.600 &  19.397 &  13.0 &    DQ\\
221528.01$+$065220.7 &  2308-54379-0592 &   5.313 &  18.362 &  18.604 &  17.8 &    dC\\
221850.23$+$212303.7 &  5946-56101-0356 &  10.364 &  17.929 &  17.902 &  29.9 &    DQ\\
222032.07$+$142032.5 &  5042-55856-0642 &   7.431 &  18.621 &  18.582 &  30.5 &    DQ\\
222552.64$+$125116.8 &  5042-55856-0096 &   6.426 &  19.774 &  19.594 &  18.8 &    DQ\\
223224.00$-$074434.3 &  0721-52228-0533 &  18.268 &  18.413 &  17.912 &  25.8 &    DQpec\\
225459.62$+$002545.3 &  4206-55471-0156 &   5.483 &  19.285 &  19.225 &  20.5 &    DQ\\
225901.16$+$215843.9 &  6591-56535-0384 &   6.805 &  20.751 &  20.226 &   6.3 &    DQpec\\
230249.36$+$243027.1 &  6588-56536-0038 &  13.792 &  18.177 &  17.971 &  28.9 &    DQ\\
231030.24$+$005745.6 &  4209-55478-0404 &   6.949 &  19.153 &  19.045 &  18.9 &    DQ\\
234130.75$+$151943.5 &  1894-53240-0463 &  13.082 &  18.347 &  17.458 &  19.3 &    dC\\
234132.83$-$010104.5 &  0385-51877-0126 &   8.158 &  18.309 &  18.233 &  20.1 &    DQ\\
234843.30$-$094245.4 &  7166-56602-0536 &   2.764 &  19.028 &  19.263 &  15.5 &    hDQ\\
235443.13$+$362907.2 &  1880-53262-0153 &  13.082 &  17.599 &  17.458 &  27.8 &    dC\\
\end{longtable}
}

\longtab[2]{
\begin{longtable}{cccccc}
\caption{\label{tabfitdq} Fit results for the classical DQ with Swan
  bands. The numbers in parentheses for \hbox{$T\sb{\rm eff}$}, \hbox{$\log g$},
  [C/He], and $M$/\hbox{M$\sb{\odot}$}\ are the internal uncertainties. }
\\ \hline\hline SDSS~J & p-m-f & Teff[K] & log g [cgs] & [C/He] &
$M$/\hbox{M$\sb{\odot}$} \\ \hline \endfirsthead
\caption{continued.}\\
\hline\hline
  SDSS~J   &     p-m-f   &      Teff[K] & log g [cgs] &  [C/He]   &
  $M$/\hbox{M$\sb{\odot}$} \\
\hline
\endhead
\hline
\endfoot               % 221 DQ objects
0000$-$0850 & 7167-56604-0752 & 7944 (117)& 7.453 (0.164)& $-5.172$ (1.369)& 0.311 (0.060)\\
0007$+$2821 & 2824-54452-0602 & 7529 ( 92)& 7.931 (0.142)& $-5.427$ (0.584)& 0.536 (0.082)\\
0015$+$0309 & 4298-55511-0906 & 7256 (132)& 8.004 (0.181)& $-5.678$ (2.117)& 0.578 (0.109)\\
0025$-$1108 & 0653-52145-0086 & 8402 ( 82)& 7.866 (0.041)& $-4.754$ (0.291)& 0.502 (0.023)\\
0033$+$0418 & 4303-55508-0562 & 7700 (108)& 7.847 (0.073)& $-5.792$ (0.601)& 0.489 (0.040)\\
0044$+$1259 & 6202-56266-0820 & 7233 (128)& 8.063 (0.164)& $-5.907$ (0.754)& 0.615 (0.102)\\
0055$+$0850 & 4544-55855-0064 & 6519 (104)& 7.890 (0.061)& $-6.675$ (1.446)& 0.511 (0.034)\\
0107$+$0102 & 2328-53728-0444 & 6486 ( 57)& 7.947 (0.046)& $-6.734$ (1.172)& 0.544 (0.027)\\
0116$+$2346 & 5693-56246-0401 & 8190 ( 72)& 7.955 (0.030)& $-5.096$ (1.387)& 0.552 (0.018)\\
0127$+$0046 & 0399-51817-0099 & 8342 (101)& 7.949 (0.079)& $-4.932$ (0.918)& 0.549 (0.047)\\
0143$+$1154 & 4662-55590-0818 & 6666 (130)& 7.995 (0.339)& $-6.278$ (0.537)& 0.573 (0.200)\\
0154$+$0040 & 0403-51871-0268 & 7356 ( 75)& 7.934 (0.052)& $-5.700$ (0.492)& 0.537 (0.030)\\
0154$+$1403 & 0430-51877-0558 & 6446 ( 49)& 7.898 (0.028)& $-6.697$ (0.720)& 0.515 (0.016)\\
0200$+$0714 & 4531-55563-0089 & 6772 ( 77)& 8.008 (0.047)& $-6.097$ (1.343)& 0.580 (0.029)\\
0205$+$2156 & 7636-56989-0171 & 5984 ( 57)& 8.281 (0.083)& $-6.277$ (0.232)& 0.754 (0.056)\\
0229$+$2510 & 2399-53764-0244 & 7979 (109)& 7.975 (0.067)& $-5.087$ (1.716)& 0.563 (0.040)\\
0239$+$0027 & 3745-55234-0924 & 7255 (101)& 7.635 (0.202)& $-5.586$ (1.691)& 0.383 (0.091)\\
0248$+$3408 & 2398-53768-0283 & 5786 ( 41)& 7.850 (0.038)& $-7.050$ (0.497)& 0.486 (0.021)\\
0253$+$3414 & 2398-53768-0154 & 7647 (111)& 7.475 (0.303)& $-5.730$ (2.309)& 0.318 (0.115)\\
0320$-$0716 & 0460-51924-0236 & 6819 ( 63)& 8.114 (0.078)& $-4.468$ (0.117)& 0.647 (0.050)\\
0332$+$0037 & 2069-53376-0148 & 8285 ( 83)& 8.042 (0.041)& $-4.903$ (0.395)& 0.605 (0.025)\\
0416$+$0713 & 2826-54389-0369 & 7750 (116)& 8.016 (0.115)& $-5.421$ (0.759)& 0.587 (0.070)\\
0723$+$3908 & 3657-55244-0476 & 9339 (110)& 7.898 (0.042)& $-4.479$ (0.453)& 0.520 (0.024)\\
0731$+$6636 & 2944-54523-0333 & 8863 (127)& 7.985 (0.110)& $-4.575$ (1.300)& 0.570 (0.066)\\
0737$+$6455 & 2944-54523-0244 & 7379 ( 83)& 7.740 (0.149)& $-5.440$ (0.449)& 0.433 (0.074)\\
0740$+$1810 & 2915-54497-0274 & 6991 ( 83)& 8.128 (0.124)& $-5.931$ (0.479)& 0.656 (0.080)\\
0742$+$2411 & 4470-55587-0840 & 8385 ( 91)& 7.926 (0.114)& $-4.902$ (0.570)& 0.536 (0.065)\\
0742$+$4348 & 3669-55481-0722 & 8119 (101)& 7.979 (0.059)& $-4.986$ (0.294)& 0.566 (0.035)\\
0750$+$1328 & 4502-55569-0218 & 8144 ( 85)& 7.910 (0.073)& $-5.239$ (1.299)& 0.526 (0.041)\\
0750$+$2329 & 2916-54507-0474 & 7465 (141)& 8.239 (0.175)& $-5.301$ (0.448)& 0.730 (0.115)\\
0752$+$3019 & 3752-55236-0652 & 7194 (135)& 8.254 (0.185)& $-5.957$ (0.912)& 0.739 (0.122)\\
0752$+$4447 & 3671-55483-0028 & 8745 ( 72)& 7.863 (0.022)& $-4.904$ (0.434)& 0.500 (0.012)\\
0802$+$4147 & 3683-55178-0162 & 7276 (133)& 7.872 (0.292)& $-5.807$ (0.694)& 0.502 (0.161)\\
0804$+$0750 & 2076-53442-0408 & 7921 (111)& 7.932 (0.230)& $-5.252$ (0.690)& 0.538 (0.132)\\
0808$+$4640 & 0438-51884-0063 & 5155 ( 29)& 7.597 (0.123)& $-5.998$ (0.153)& 0.358 (0.056)\\
0813$+$3047 & 0861-52318-0053 & 7720 ( 85)& 7.966 (0.070)& $-5.254$ (0.432)& 0.557 (0.041)\\
0814$+$2455 & 4463-55868-0140 & 7996 ( 64)& 7.950 (0.022)& $-5.418$ (0.368)& 0.548 (0.013)\\
0822$+$2023 & 1927-53321-0125 & 6987 ( 49)& 7.553 (0.023)& $-6.849$ (0.783)& 0.346 (0.009)\\
0827$+$2100 & 4483-55587-0850 & 6829 ( 87)& 7.887 (0.087)& $-6.312$ (0.524)& 0.509 (0.049)\\
0833$+$3638 & 0864-52320-0314 & 7777 (101)& 7.976 (0.086)& $-5.460$ (0.996)& 0.563 (0.051)\\
0836$+$0437 & 3809-55533-0504 & 8588 (116)& 7.840 (0.090)& $-4.743$ (0.588)& 0.488 (0.049)\\
0836$+$4817 & 0550-51959-0433 & 7304 ( 72)& 7.900 (0.063)& $-6.051$ (5.108)& 0.518 (0.036)\\
0837$+$0321 & 3809-55533-0466 & 8223 (125)& 7.951 (0.111)& $-5.399$ (1.136)& 0.550 (0.065)\\
0838$+$1121 & 5284-55866-0786 & 7449 ( 92)& 8.052 (0.088)& $-5.411$ (0.353)& 0.609 (0.054)\\
0839$+$0842 & 5284-55866-0206 & 6134 ( 84)& 8.021 (0.133)& $-6.251$ (0.205)& 0.586 (0.081)\\
0840$+$4529 & 5163-55889-0646 & 7883 ( 73)& 7.997 (0.040)& $-5.267$ (0.255)& 0.576 (0.024)\\
0841$+$3329 & 0933-52642-0016 & 6912 ( 58)& 8.711 (0.058)& $-6.165$ (1.355)& 1.027 (0.033)\\
0845$+$6143 & 1875-54453-0184 & 8116 ( 63)& 7.978 (0.023)& $-5.008$ (0.851)& 0.565 (0.014)\\
0846$+$1024 & 2671-54141-0310 & 9197 (147)& 7.825 (0.058)& $-4.625$ (0.542)& 0.480 (0.031)\\
0850$+$0709 & 5289-55893-0088 & 8637 (125)& 7.623 (0.165)& $-4.929$ (0.709)& 0.382 (0.072)\\
0855$+$0639 & 1189-52668-0535 & 7359 ( 71)& 7.962 (0.034)& $-5.750$ (0.378)& 0.554 (0.020)\\
0857$+$0603 & 1189-52668-0027 & 8418 ( 95)& 8.000 (0.051)& $-4.751$ (0.466)& 0.579 (0.031)\\
0902$+$5037 & 0551-51993-0612 & 7703 (118)& 7.901 (0.113)& $-5.721$ (2.718)& 0.519 (0.064)\\
0904$+$3954 & 1199-52703-0595 & 7416 ( 98)& 8.091 (0.123)& $-5.364$ (0.597)& 0.633 (0.078)\\
0905$+$0904 & 5299-55927-0646 & 8254 ( 59)& 7.863 (0.019)& $-5.076$ (0.745)& 0.500 (0.010)\\
0906$+$4702 & 0898-52606-0565 & 5472 ( 52)& 8.209 (0.060)& $-5.087$ (0.306)& 0.706 (0.040)\\
0915$+$2019 & 5768-56017-0666 & 8791 (133)& 7.971 (0.158)& $-4.689$ (1.524)& 0.562 (0.094)\\
0916$+$1011 & 5302-55896-0023 & 8451 ( 59)& 7.925 (0.017)& $-4.866$ (0.212)& 0.535 (0.010)\\
0920$+$3603 & 4644-55922-0898 & 7841 ( 98)& 7.991 (0.094)& $-5.434$ (0.663)& 0.572 (0.056)\\
0920$+$5306 & 7289-57039-0974 & 8948 (121)& 8.212 (0.186)& $-3.417$ (0.305)& 0.714 (0.121)\\
0921$+$3421 & 5810-56358-0620 & 8260 ( 82)& 7.436 (0.103)& $-4.865$ (0.398)& 0.307 (0.036)\\
0926$+$4725 & 0900-52637-0041 & 7175 ( 56)& 7.975 (0.042)& $-6.165$ (1.041)& 0.561 (0.025)\\
0929$+$3310 & 1593-52991-0094 & 6659 ( 36)& 7.909 (0.053)& $-4.655$ (0.080)& 0.522 (0.031)\\
0930$+$2959 & 2914-54533-0304 & 8288 ( 93)& 7.819 (0.134)& $-4.911$ (0.364)& 0.476 (0.071)\\
0934$+$1158 & 5313-55973-0175 & 8482 (124)& 7.827 (0.098)& $-4.962$ (0.707)& 0.481 (0.052)\\
0935$+$2417 & 2294-53733-0044 & 8729 (101)& 7.930 (0.068)& $-4.669$ (1.280)& 0.538 (0.040)\\
0939$+$5201 & 5724-56364-0374 & 8473 ( 48)& 7.949 (0.019)& $-4.875$ (0.321)& 0.549 (0.011)\\
0940$+$0210 & 0477-52026-0493 & 7256 ( 62)& 8.094 (0.022)& $-5.769$ (0.186)& 0.634 (0.014)\\
0940$+$3843 & 3224-54849-0517 & 7183 ( 79)& 7.537 (0.099)& $-6.244$ (1.100)& 0.341 (0.041)\\
0941$+$0901 & 1304-52993-0009 & 8608 ( 54)& 7.344 (0.025)& $-4.903$ (0.499)& 0.278 (0.008)\\
0941$+$4414 & 4695-55957-0208 & 8266 ( 85)& 8.003 (0.056)& $-4.940$ (0.544)& 0.581 (0.034)\\
0944$+$5827 & 5715-56657-0715 & 8995 (106)& 7.953 (0.066)& $-4.462$ (0.697)& 0.551 (0.039)\\
0948$+$1232 & 1742-53053-0616 & 7195 ( 91)& 8.124 (0.040)& $-6.187$ (1.134)& 0.653 (0.026)\\
0950$+$5315 & 0769-52282-0135 & 7993 ( 46)& 7.884 (0.015)& $-5.373$ (0.343)& 0.510 (0.009)\\
0951$+$6243 & 2403-53795-0577 & 8706 (144)& 7.886 (0.090)& $-4.747$ (0.903)& 0.513 (0.051)\\
1005$-$0114 & 3769-55240-0546 & 8519 (128)& 7.930 (0.186)& $-4.801$ (0.604)& 0.538 (0.107)\\
1010$+$2300 & 6458-56274-0618 & 8591 (109)& 7.921 (0.073)& $-4.915$ (0.561)& 0.533 (0.042)\\
1012$+$0040 & 0502-51957-0095 & 9327 ( 85)& 7.802 (0.051)& $-4.330$ (0.590)& 0.468 (0.026)\\
1013$+$4350 & 4561-55614-0788 & 8507 (124)& 7.946 (0.073)& $-4.906$ (0.767)& 0.547 (0.043)\\
1015$+$3518 & 4568-55600-0256 & 8186 ( 97)& 7.804 (0.128)& $-4.997$ (2.123)& 0.468 (0.067)\\
1017$+$3736 & 1427-52996-0216 & 7167 ( 68)& 7.944 (0.077)& $-5.835$ (0.454)& 0.542 (0.045)\\
1018$+$0838 & 1237-52762-0621 & 7644 ( 61)& 7.928 (0.022)& $-5.764$ (0.316)& 0.535 (0.013)\\
1022$+$2845 & 2351-53772-0170 & 7221 (101)& 7.945 (0.082)& $-5.800$ (0.498)& 0.543 (0.048)\\
1026$+$2725 & 6463-56340-0894 & 8342 (111)& 7.970 (0.106)& $-4.937$ (0.610)& 0.561 (0.063)\\
1027$+$1218 & 1747-53075-0241 & 6684 ( 58)& 7.945 (0.037)& $-6.731$ (1.383)& 0.543 (0.048)\\
1028$+$2537 & 2349-53734-0535 & 6603 ( 68)& 7.872 (0.051)& $-6.470$ (0.457)& 0.501 (0.029)\\
1028$+$3512 & 1958-53385-0558 & 5627 ( 26)& 8.004 (0.030)& $-5.667$ (0.048)& 0.574 (0.019)\\
1031$+$2217 & 5873-56035-0828 & 7070 (103)& 8.601 (0.159)& $-5.857$ (0.614)& 0.964 (0.095)\\
1032$+$2101 & 5873-56035-0852 & 8297 (107)& 8.030 (0.084)& $-4.909$ (0.623)& 0.597 (0.052)\\
1032$+$4519 & 4691-55651-0758 & 9116 ( 99)& 7.933 (0.053)& $-4.435$ (0.392)& 0.540 (0.031)\\
1040$+$1822 & 5886-56034-0578 & 6812 (146)& 8.027 (0.197)& $-6.148$ (1.863)& 0.592 (0.120)\\
1042$+$5833 & 7095-56625-0462 & 7312 ( 82)& 8.071 (0.065)& $-5.620$ (1.534)& 0.620 (0.041)\\
1045$+$2134 & 5874-56039-0954 & 7768 (104)& 7.928 (0.098)& $-5.292$ (0.536)& 0.535 (0.056)\\
1052$+$5911 & 7095-56625-0848 & 7920 ( 90)& 7.906 (0.042)& $-5.371$ (0.371)& 0.523 (0.024)\\
1058$+$3440 & 4631-55617-0148 & 6737 (111)& 8.150 (0.159)& $-6.573$ (9.735)& 0.670 (0.102)\\
1101$+$5218 & 6706-56385-0856 & 8591 (121)& 7.943 (0.105)& $-4.731$ (0.498)& 0.546 (0.061)\\
1108$+$1349 & 5360-55983-0034 & 7616 ( 86)& 8.019 (0.038)& $-5.879$ (0.505)& 0.589 (0.023)\\
1113$+$4455 & 6649-56364-0050 & 8049 (108)& 7.940 (0.115)& $-5.201$ (1.100)& 0.543 (0.067)\\
1120$+$4252 & 1440-53084-0408 & 7476 ( 81)& 8.018 (0.058)& $-5.903$ (1.263)& 0.588 (0.036)\\
1123$+$3347 & 4619-55599-0440 & 7903 ( 94)& 8.191 (0.129)& $-5.317$ (0.614)& 0.699 (0.084)\\
1123$+$3523 & 4618-55600-0334 & 6751 (107)& 7.856 (0.264)& $-6.037$ (3.413)& 0.492 (0.144)\\
1123$+$6207 & 3328-54964-0497 & 7050 ( 74)& 7.761 (0.088)& $-5.388$ (0.263)& 0.442 (0.045)\\
1126$+$4419 & 3215-54861-0257 & 6943 ( 77)& 7.989 (0.064)& $-6.096$ (1.374)& 0.569 (0.038)\\
1131$+$1845 & 5879-56047-0298 & 8515 (127)& 7.938 (0.147)& $-5.072$ (2.902)& 0.543 (0.085)\\
1133$+$1900 & 5879-56047-0178 & 7906 (101)& 7.700 (0.147)& $-5.256$ (0.374)& 0.415 (0.070)\\
1135$+$5724 & 1310-53033-0485 & 7338 ( 49)& 7.333 (0.021)& $-6.137$ (0.874)& 0.267 (0.007)\\
1140$+$1540 & 1755-53386-0607 & 6365 ( 68)& 7.859 (0.080)& $-6.600$ (0.694)& 0.493 (0.044)\\
1142$+$0352 & 4766-55677-0262 & 7742 (132)& 7.731 (0.140)& $-5.302$ (0.459)& 0.429 (0.069)\\
1151$+$4527 & 6642-56396-0596 & 8756 ( 96)& 7.911 (0.036)& $-4.556$ (1.553)& 0.527 (0.021)\\
1200$+$4237 & 1447-53120-0621 & 6287 ( 70)& 8.169 (0.049)& $-6.456$ (0.659)& 0.681 (0.032)\\
1201$+$3400 & 2089-53498-0318 & 5953 ( 27)& 8.117 (0.015)& $-5.947$ (0.084)& 0.646 (0.009)\\
1211$+$0440 & 4749-55633-0722 & 7745 (114)& 7.939 (0.095)& $-5.540$ (4.841)& 0.541 (0.055)\\
1212$+$5452 & 6688-56412-0854 & 8203 ( 89)& 7.963 (0.031)& $-4.994$ (0.389)& 0.617 (0.048)\\
1222$+$6041 & 6968-56443-0588 & 9155 (151)& 8.061 (0.076)& $-4.407$ (0.653)& 0.617 (0.047)\\
1225$+$4706 & 1451-53117-0035 & 6241 ( 54)& 7.913 (0.067)& $-4.999$ (0.367)& 0.523 (0.038)\\
1228$+$4806 & 6671-56388-0004 & 6948 (114)& 7.806 (0.112)& $-5.888$ (0.684)& 0.465 (0.059)\\
1235$+$3918 & 3970-55591-0354 & 9347 ( 89)& 8.041 (0.024)& $-4.107$ (0.352)& 0.605 (0.015)\\
1240$-$0144 & 2922-54612-0392 & 8545 (125)& 7.878 (0.160)& $-4.735$ (0.458)& 0.508 (0.089)\\
1243$+$3607 & 2022-53827-0238 & 8531 (117)& 7.956 (0.047)& $-4.642$ (1.497)& 0.553 (0.027)\\
1244$+$2709 & 2238-54205-0229 & 7959 (108)& 7.273 (0.069)& $-4.865$ (0.369)& 0.252 (0.021)\\
1247$+$4113 & 4703-55617-0438 & 8596 (112)& 7.998 (0.038)& $-4.867$ (0.463)& 0.578 (0.023)\\
1247$+$4915 & 2923-54563-0222 & 5934 ( 56)& 7.800 (0.061)& $-7.082$ (1.232)& 0.460 (0.032)\\
1249$+$3407 & 3971-55322-0018 & 8466 ( 82)& 7.942 (0.026)& $-4.889$ (0.310)& 0.545 (0.015)\\
1251$+$4646 & 1457-53116-0215 & 8687 ( 84)& 7.898 (0.032)& $-4.883$ (1.016)& 0.520 (0.018)\\
1252$+$1943 & 2615-54483-0381 & 6779 ( 57)& 7.536 (0.060)& $-6.365$ (0.309)& 0.338 (0.025)\\
1253$+$0139 & 0523-52026-0252 & 8650 (138)& 7.962 (0.105)& $-4.619$ (2.323)& 0.557 (0.062)\\
1302$+$0923 & 3234-54885-0143 & 7342 ( 79)& 7.924 (0.054)& $-5.552$ (0.769)& 0.531 (0.031)\\
1315$+$4711 & 6625-56386-0556 & 7630 ( 73)& 7.861 (0.028)& $-5.747$ (0.569)& 0.497 (0.015)\\
1316$+$0810 & 5423-55958-0996 & 6713 ( 54)& 7.992 (0.041)& $-6.241$ (1.479)& 0.571 (0.024)\\
1319$+$0844 & 5429-55979-0598 & 8485 ( 92)& 7.953 (0.041)& $-5.033$ (3.097)& 0.551 (0.024)\\
1319$+$1401 & 5425-56003-0148 & 7797 ( 98)& 7.922 (0.071)& $-5.212$ (0.634)& 0.531 (0.041)\\
1322$+$3730 & 2102-54115-0356 & 7732 (110)& 8.016 (0.065)& $-5.552$ (4.268)& 0.587 (0.040)\\
1328$+$3640 & 3983-55603-0800 & 8285 ( 88)& 7.967 (0.043)& $-4.808$ (0.246)& 0.559 (0.025)\\
1329$+$0746 & 1801-54156-0324 & 8163 ( 64)& 8.005 (0.022)& $-5.280$ (0.414)& 0.582 (0.013)\\  
1331$+$6704 & 6825-56717-0629 & 9011 (135)& 7.984 (0.063)& $-4.465$ (0.779)& 0.570 (0.038)\\
1333$+$2357 & 5995-56093-0958 & 7998 (100)& 8.029 (0.058)& $-5.172$ (2.095)& 0.596 (0.036)\\
1334$+$1622 & 5434-56033-0075 & 8859 (114)& 7.993 (0.042)& $-4.644$ (0.590)& 0.575 (0.025)\\
1336$+$1920 & 5862-56045-0694 & 6199 ( 72)& 8.026 (0.094)& $-6.748$ (0.763)& 0.590 (0.058)\\
1339$+$5014 & 6744-56399-0626 & 8739 ( 85)& 7.952 (0.043)& $-4.867$ (0.997)& 0.551 (0.025)\\
1344$+$1849 & 2930-54589-0391 & 8243 ( 95)& 7.819 (0.043)& $-5.751$ (1.954)& 0.476 (0.023)\\
1347$+$1528 & 5441-56017-0848 & 8451 (125)& 8.228 (0.080)& $-4.750$ (0.419)& 0.724 (0.052)\\
1347$+$3817 & 3852-55243-0630 & 8182 (114)& 8.119 (0.073)& $-5.169$ (1.607)& 0.653 (0.047)\\
1347$+$5019 & 6740-56401-0202 & 8437 ( 66)& 7.978 (0.068)& $-3.827$ (0.145)& 0.566 (0.041)\\
1351$+$6623 & 6986-56717-0616 & 8985 ( 84)& 7.975 (0.025)& $-4.559$ (3.099)& 0.564 (0.015)\\
1352$+$2218 & 5869-56064-0716 & 8068 (104)& 7.970 (0.093)& $-4.972$ (0.522)& 0.560 (0.055)\\
1352$+$2658 & 6005-56090-0670 & 7939 (273)& 8.000 (0.103)& $-5.259$ (0.588)& 0.578 (0.062)\\
1354$+$1217 & 5443-56010-0064 & 7845 (114)& 7.626 (0.237)& $-5.437$ (0.753)& 0.380 (0.105)\\
1355$+$3636 & 3852-55243-0048 & 8219 ( 72)& 7.971 (0.026)& $-4.324$ (0.091)& 0.561 (0.015)\\
1356$+$0009 & 0301-51942-0231 & 6587 ( 37)& 8.073 (0.030)& $-5.491$ (0.256)& 0.621 (0.019)\\
1357$+$2949 & 2122-54178-0346 & 7624 ( 81)& 7.945 (0.047)& $-5.473$ (0.513)& 0.544 (0.027)\\
1358$+$0552 & 4863-55688-0462 & 8401 (123)& 7.904 (0.108)& $-4.867$ (0.611)& 0.523 (0.062)\\
1406$+$0148 & 0532-51993-0234 & 7794 ( 88)& 7.923 (0.050)& $-5.464$ (0.759)& 0.532 (0.029)\\
1406$+$0204 & 0532-51993-0475 & 7721 ( 90)& 7.357 (0.085)& $-5.477$ (1.020)& 0.277 (0.028)\\
1407$+$2039 & 5894-56039-0782 & 8320 ( 98)& 7.724 (0.095)& $-4.886$ (0.519)& 0.428 (0.047)\\
1416$+$3016 & 6497-56329-0888 & 7595 (158)& 8.205 (0.262)& $-5.661$ (1.586)& 0.707 (0.169)\\
1417$+$2412 & 2128-53800-0153 & 8129 ( 76)& 8.105 (0.046)& $-4.891$ (0.444)& 0.644 (0.030)\\
1420$+$5316 & 7028-56449-0881 & 7392 (101)& 7.941 (0.074)& $-5.852$ (0.803)& 0.541 (0.043)\\
1424$+$0833 & 5462-55978-0432 & 9329 (110)& 7.891 (0.041)& $-4.430$ (0.437)& 0.516 (0.023)\\
1425$+$1801 & 5897-56042-0259 & 7993 (116)& 7.894 (0.067)& $-5.146$ (1.308)& 0.516 (0.038)\\
1427$+$6110 & 0607-52368-0379 & 6313 ( 39)& 7.894 (0.018)& $-6.666$ (0.249)& 0.513 (0.010)\\
1431$+$3750 & 1381-53089-0599 & 6313 ( 78)& 7.358 (0.104)& $-6.727$ (0.674)& 0.270 (0.036)\\
1432$+$4751 & 6752-56366-0851 & 8563 ( 92)& 7.908 (0.037)& $-4.934$ (0.618)& 0.525 (0.021)\\
1435$+$2120 & 5900-56043-0959 & 8780 (157)& 7.947 (0.132)& $-4.581$ (2.941)& 0.548 (0.077)\\
1442$+$4202 & 6061-56076-0860 & 8812 (129)& 7.964 (0.104)& $-4.713$ (0.975)& 0.558 (0.062)\\
1445$+$1851 & 5903-56064-0342 & 8195 (113)& 7.802 (0.108)& $-5.120$ (1.992)& 0.467 (0.057)\\  
1448$+$0047 & 0308-51662-0145 & 7075 ( 63)& 8.011 (0.041)& $-6.127$ (1.491)& 0.582 (0.025)\\
1449$+$2916 & 3878-55361-0654 & 7966 ( 97)& 8.011 (0.094)& $-5.190$ (1.462)& 0.585 (0.057)\\
1450$+$6350 & 2947-54533-0605 & 8060 (137)& 7.799 (0.188)& $-4.928$ (0.838)& 0.465 (0.098)\\
1501$+$0627 & 1815-53884-0095 & 7216 ( 74)& 8.167 (0.064)& $-5.755$ (0.836)& 0.681 (0.042)\\
1501$+$2100 & 2150-54510-0135 & 5728 ( 27)& 7.228 (0.046)& $-6.140$ (0.094)& 0.223 (0.014)\\
1509$+$1620 & 5484-56039-0232 & 8398 (113)& 7.871 (0.131)& $-5.132$ (2.098)& 0.504 (0.072)\\
1511$+$5008 & 1330-52822-0417 & 7836 ( 98)& 7.973 (0.056)& $-5.466$ (1.095)& 0.561 (0.034)\\
1514$+$1423 & 2766-54242-0126 & 7094 ( 40)& 8.296 (0.054)& $-3.998$ (0.330)& 0.767 (0.036)\\
1517$+$2256 & 3955-55678-0592 & 7530 ( 61)& 7.983 (0.025)& $-5.752$ (0.277)& 0.567 (0.015)\\
1519$+$3856 & 4979-56045-0654 & 8293 (111)& 7.926 (0.067)& $-4.974$ (0.680)& 0.535 (0.038)\\
1520$+$2703 & 3851-55302-0214 & 7849 (113)& 7.808 (0.100)& $-5.368$ (0.408)& 0.469 (0.052)\\
1528$+$0442 & 1835-54563-0179 & 7485 ( 83)& 7.925 (0.110)& $-5.685$ (3.812)& 0.532 (0.063)\\
1528$+$4003 & 2936-54626-0548 & 7734 (108)& 8.061 (0.056)& $-5.478$ (0.554)& 0.615 (0.035)\\
1528$+$5134 & 6721-56398-0969 & 7626 ( 66)& 8.020 (0.030)& $-5.356$ (0.268)& 0.589 (0.019)\\
1534$+$4145 & 1679-53149-0616 & 7745 ( 76)& 7.907 (0.026)& $-5.794$ (0.572)& 0.523 (0.015)\\
1537$+$1337 & 4890-55741-0439 & 8017 ( 83)& 7.896 (0.055)& $-5.348$ (0.548)& 0.517 (0.031)\\
1542$+$1314 & 4890-55741-0172 & 8150 ( 83)& 7.994 (0.047)& $-5.153$ (1.412)& 0.575 (0.029)\\
1542$+$2544 & 3947-55332-0760 & 6691 ( 64)& 7.990 (0.059)& $-6.292$ (0.392)& 0.570 (0.036)\\
1548$+$5626 & 0617-52072-0551 & 8013 (108)& 8.066 (0.054)& $-5.274$ (1.012)& 0.619 (0.034)\\
1551$+$0824 & 1727-53859-0442 & 6741 ( 58)& 7.932 (0.029)& $-6.146$ (0.979)& 0.535 (0.017)\\
1552$+$1148 & 2520-54584-0515 & 7870 (110)& 8.009 (0.076)& $-5.216$ (0.751)& 0.583 (0.046)\\
1552$+$3910 & 5192-56066-0472 & 7614 ( 81)& 8.055 (0.040)& $-5.524$ (2.216)& 0.611 (0.025)\\
1554$+$0336 & 0595-52023-0373 & 6716 ( 75)& 7.893 (0.053)& $-6.411$ (0.674)& 0.513 (0.030)\\
1554$+$1136 & 2520-54584-0630 & 8382 (107)& 7.865 (0.087)& $-4.750$ (0.388)& 0.501 (0.048)\\
1609$+$0655 & 1730-53498-0217 & 8012 ( 98)& 7.934 (0.036)& $-5.438$ (0.713)& 0.539 (0.021)\\
1610$+$3036 & 5009-55707-0838 & 8133 ( 96)& 7.986 (0.063)& $-4.925$ (0.455)& 0.570 (0.038)\\
1613$+$5116 & 6316-56483-0713 & 7990 (120)& 8.011 (0.072)& $-5.077$ (2.317)& 0.585 (0.044)\\
1616$+$3924 & 1336-52759-0572 & 7257 ( 60)& 7.955 (0.028)& $-5.787$ (0.384)& 0.549 (0.016)\\
1620$+$1809 & 4060-55359-0444 & 7859 ( 68)& 7.889 (0.034)& $-5.433$ (0.307)& 0.513 (0.019)\\
1621$+$2253 & 4184-55450-0463 & 8147 ( 97)& 7.957 (0.045)& $-5.259$ (0.479)& 0.553 (0.026)\\
1627$+$3043 & 4953-55749-0732 & 8406 ( 88)& 7.870 (0.058)& $-5.130$ (2.155)& 0.504 (0.032)\\
1635$+$2843 & 5201-55832-0412 & 7890 (105)& 8.188 (0.061)& $-5.244$ (1.038)& 0.697 (0.040)\\
1638$+$1244 & 4075-55352-0802 & 8669 ( 79)& 7.933 (0.032)& $-4.805$ (0.337)& 0.540 (0.019)\\
1641$+$4833 & 6318-56186-0724 & 7735 ( 75)& 8.021 (0.028)& $-5.476$ (0.389)& 0.590 (0.017)\\
1654$+$3157 & 1176-52791-0238 & 7389 ( 63)& 8.033 (0.022)& $-5.479$ (0.264)& 0.597 (0.014)\\
1655$+$3722 & 5198-55823-0844 & 8881 ( 83)& 7.928 (0.045)& $-4.436$ (0.533)& 0.537 (0.026)\\
1712$+$3406 & 4994-55739-0426 & 8145 (106)& 7.954 (0.084)& $-5.148$ (2.366)& 0.551 (0.049)\\
1713$+$3240 & 4998-55722-0536 & 8070 ( 55)& 7.999 (0.018)& $-4.970$ (0.191)& 0.578 (0.011)\\
2046$-$0715 & 0635-52145-0228 & 8147 ( 98)& 7.769 (0.075)& $-5.226$ (2.019)& 0.449 (0.038)\\
2053$-$0702 & 0636-52176-0267 & 6726 ( 38)& 8.075 (0.049)& $-4.648$ (0.175)& 0.622 (0.031)\\
2135$+$0003 & 0989-52468-0198 & 6379 ( 56)& 7.865 (0.202)& $-6.746$ (0.935)& 0.497 (0.112)\\
2140$+$0045 & 4196-55478-0444 & 7605 (100)& 8.039 (0.102)& $-5.495$ (0.412)& 0.601 (0.063)\\
2142$+$0908 & 4093-55475-0084 & 7328 (102)& 8.147 (0.173)& $-5.618$ (0.595)& 0.669 (0.112)\\
2149$+$2039 & 5949-56096-0402 & 8873 (118)& 7.684 (0.185)& $-4.182$ (0.150)& 0.409 (0.086)\\
2150$-$0113 & 4197-55479-0362 & 9046 (103)& 7.962 (0.038)& $-4.542$ (0.727)& 0.557 (0.023)\\
2156$+$0559 & 4096-55501-0804 & 6973 ( 87)& 8.035 (0.047)& $-6.249$ (1.361)& 0.597 (0.029)\\
2157$+$1137 & 5063-55831-0387 & 8711 (101)& 7.839 (0.079)& $-4.632$ (1.019)& 0.487 (0.043)\\
2207$+$0135 & 4316-55505-0170 & 6432 ( 73)& 8.067 (0.073)& $-6.698$ (1.009)& 0.617 (0.045)\\
2212$+$0613 & 2323-54380-0113 & 6730 ( 84)& 8.280 (0.104)& $-6.174$ (1.164)& 0.756 (0.069)\\
2218$+$2123 & 5946-56101-0356 & 8255 ( 62)& 7.983 (0.029)& $-4.951$ (0.358)& 0.569 (0.018)\\
2220$+$1420 & 5042-55856-0642 & 8670 (122)& 8.050 (0.070)& $-4.750$ (0.569)& 0.610 (0.043)\\
2225$+$1251 & 5042-55856-0096 & 6498 (130)& 7.851 (0.216)& $-6.519$ (1.670)& 0.489 (0.118)\\
2254$+$0025 & 4206-55471-0156 & 8295 (130)& 7.867 (0.142)& $-4.967$ (0.458)& 0.502 (0.079)\\
2302$+$2430 & 6588-56536-0038 & 6863 ( 53)& 8.017 (0.026)& $-6.357$ (0.264)& 0.586 (0.016)\\
2310$+$0057 & 4209-55478-0404 & 7881 ( 67)& 8.017 (0.102)& $-3.971$ (0.064)& 0.588 (0.062)\\
2341$-$0101 & 0385-51877-0126 & 8568 (117)& 7.924 (0.059)& $-4.897$ (0.614)& 0.535 (0.034)\\
\end{longtable}
}
\end{appendix}

\end{document}